\documentclass[aps,pre,twocolumn,showpacs,superscriptaddress,longbibliography]{revtex4-1}
\usepackage{amsmath} \usepackage{amssymb} 
\usepackage{graphicx} 
\usepackage{color}
\newcommand{\LLLC}{Doctoral College for the Statistical Physics of Complex Systems,
Leipzig-Lorraine-Lviv-Coventry $({\mathbb L}^4)$,\\ 
Postfach 100\,920, 04009 Leipzig, Germany}
\newcommand{\Leipzig}{Institut f\"ur Theoretische Physik, Universit\"at Leipzig Postfach 100\,920, 04009 Leipzig, Germany}
\newcommand{\Lviv}{Institute for Condensed Matter Physics of the National
Academy of Sciences of Ukraine, 79011 Lviv, Ukraine}
\newcommand{\MPIDS}{
  Max Planck Institute for Dynamics and Self-Organization, Am Fassberg 17, 37077 G{\"o}ttingen, Germany}
\newcommand{\BCCN}{
  Bernstein Center for Computational Neuroscience, Am Fassberg 17, 37077 G{\"o}ttingen, Germany
}

\begin{document}

\title{Percolation thresholds and fractal dimensions for square and cubic
lattices with long-range correlated defects}

\author{Johannes Zierenberg}
\affiliation{\Leipzig}%
\affiliation{\LLLC}%
\affiliation{\MPIDS}%
\affiliation{\BCCN}%
\author{Niklas Fricke}
\affiliation{\Leipzig}%
\affiliation{\LLLC}
\author{Martin Marenz}
\affiliation{\Leipzig}%
\affiliation{\LLLC}
\author{F.~P. Spitzner}
\affiliation{\Leipzig}
\author{Viktoria Blavatska}
\affiliation{\LLLC}%
\affiliation{\Lviv}
\author{Wolfhard Janke}
\affiliation{\Leipzig}%
\affiliation{\LLLC}

\begin{abstract}
  We study long-range power-law correlated disorder on square and cubic
  lattices. 
  In particular, we present high-precision results for the percolation
  thresholds and the fractal dimension of the largest clusters as function of
  the correlation strength. The correlations are generated using a discrete
  version of the Fourier filtering method.
  We consider two different metrics to set the length scales over which the
  correlations decay, showing that the percolation thresholds are highly
  sensitive to such system details.  By contrast, we verify that the fractal
  dimension $d_{\rm f}$ is a universal quantity and unaffected by the choice of
  metric. We also show that for weak correlations, its value coincides with
  that for the uncorrelated system. In two dimensions we observe a clear
  increase of the fractal dimension with increasing correlation strength,
  approaching $d_{\rm   f}\rightarrow 2$.  The onset of this change does not
  seem to be determined by the extended Harris criterion.

\end{abstract}

\maketitle

\affiliation{\LLLC}
\affiliation{\Leipzig}%
\renewcommand{\vec}[1]{{\bf #1}}
\renewcommand{\vr}{\vec{r}}
\newcommand{\vx}{\vec{x}}
\newcommand{\vq}{\vec{q}}
\newcommand{\vk}{\vec{k}}
\newcommand{\vl}{\vec{l}}

\newcommand{\mys}[1]{\tau_{#1}}
\newcommand{\mysd}[1]{t_{#1}}
\newcommand{\myphi}[1]{\varphi_{#1}}
\newcommand{\myphireal}[1]{\varphi^{\rm re}_{#1}}
\newcommand{\myphiimag}[1]{\varphi^{\rm im}_{#1}}
\newcommand{\myPhi}[1]{\varPhi_{#1}}
\newcommand{\myPhireal}[1]{\varPhi^{\rm re}_{#1}}
\newcommand{\myPhiimagreal}[1]{\varPhi^{\rm re / im}_{#1}}

\newcommand{\myPhiimag}[1]{\varPhi^{\rm im}_{#1}}
\newcommand{\myU}[1]    {U_{#1}}
\newcommand{\myUreal}[1]{U^{\rm re}_{#1}}
\newcommand{\myUimag}[1]{U^{\rm im}_{#1}}
\newcommand{\FTxk}[1]{e^{ \frac{2\pi i #1}{L}}}
\newcommand{\FTkx}[1]{e^{-\frac{2\pi i #1}{L}}}

\newcommand{\avgDis}[1]{\left\langle #1\right\rangle_{R}}

\newcommand{\avg}[1]{\left\langle #1\right\rangle}
\newcommand{\avgL}[1]{\left\langle #1\right\rangle_{L}}

\newcommand{\Cest}[1]{\overline{C}_{#1}}

\section{Introduction}

%

%

%

Structural obstacles (impurities) play an important role for a wide range of
physical processes as most substrates and surfaces in nature are rough and
inhomogeneous~\cite{Avnir1983,Avnir1984}. For example, the properties of
magnetic crystals are often altered by the presence of extended defects in the
form of linear dislocations or regions of different
phases~\cite{Dorogovtsev1980,Yamazaki1986a}. Another important class of such
disordered media are porous materials, which often exhibit large spatial
inhomogeneities of a fractal nature. Such fractal disorder affects a medium's
conductivity, and diffusive transport can become
anomalous~\cite{Bouchaud1990,Malek2001,Foulaadvand2015,Goychuk2017}. This aspect is
relevant, for instance, for the recovery of oil through porous
rocks~\cite{Dullien1979,Sahimi1995}, for the dynamics of fluids in disordered
media~\cite{Skinner2013, Spanner2016}, or for our understanding of transport
processes in biological cells~\cite{Bancaud2012, Hoefling-Franosch2013}.

\begin{figure}[b]
  \includegraphics{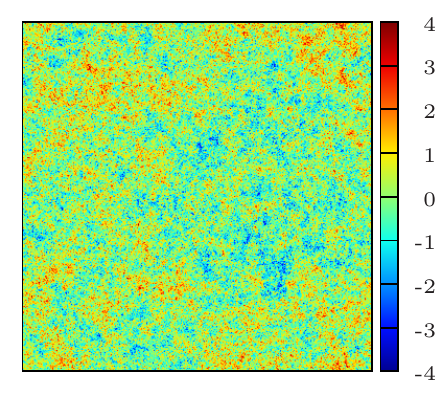}
  \hfill
  \includegraphics{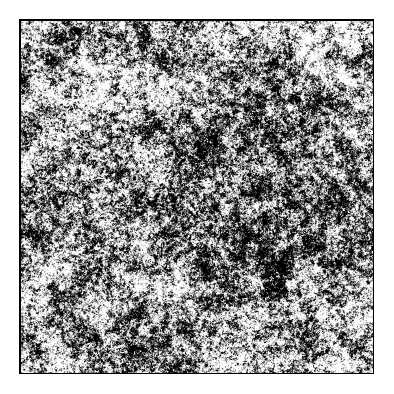}
  \caption{%
    Illustration of long-range correlated defects on a $2048^2$ lattice with
    {correlation parameter} $a=0.5$. (left) {Continuous correlated Gaussian random variables. The color reflects the value at the respective lattice site}.
    (right) Corresponding lattice of discrete variables at the finite-size
    percolation threshold $p_c^L=0.522$ with defects shown in black. 
    \label{figIllustration}
  }
\end{figure}

Disordered systems are conveniently studied in the framework of lattice
models with randomly positioned defects (or empty sites). Of particular interest is the situation
where the concentration of occupied (i.e., non-defect) lattice sites is near the
percolation threshold and clusters of connected occupied sites become fractal. The
case where defects are uncorrelated is a classic textbook model, whose
properties have been studied extensively~\cite{Stauffer1992}.  In nature,
however, inhomogeneities are often not distributed completely at random but
tend to be correlated over large distances. To understand the impact of this,
it is useful to consider the limiting case where correlations
asymptotically decay by a power law rather than exponentially with distance:
\begin{equation}\label{eqCorr}
  C(\vr)\sim |\vr|^{-a}.
\end{equation}
An illustration of such power-law correlations for continuous and discrete site
variables on a square lattice is shown in Fig.~\ref{figIllustration}. If the
{correlation parameter} $a$ is smaller than the spatial dimension $D$, the
correlations are considered long-range or ``infinite''.

The problem of power-law correlated disorder has first been investigated in the
context of spin systems and later for
percolation~\cite{Weinrib1983,Weinrib1984}. The relevance of the disorder was
shown to be characterized by an extension of the Harris criterion {for uncorrelated defects~\cite{Harris1974}}: the critical
behavior of the system deviates from the uncorrelated case if the minimum of $D$
and $a$ is smaller than $2/\nu$ (where $\nu$ denotes the correlation-length
exponent for the ordered system). It was furthermore argued that in the regime
of strong correlations, the critical correlation-length exponent for strong
disorder is universally given by $2/a$. Since $D$ is always larger than $2/\nu$
for percolation, the correlation-length exponent for long-range correlated
percolation is given by  
\begin{equation}\label{eqExtendedHarris}
\nu_a=
\begin{cases}
  2/a \quad &\text{for }  a<a_{\rm H}=2/\nu \\
  \nu \quad &\text{for }  a\geq a_{\rm H}=2/\nu. \\
\end{cases}
\end{equation}
The extended Harris criterion is still slightly
controversial~\cite{Prudnikov1999,Prudnikov2000}, but it has to some extent
been supported by numerical investigations~\cite{Schrenk2013,Prakash1992,Makse1998}.
These studies made use of the Fourier filtering method
(FFM)~\cite{Saupe1988,Peng1991,Prakash1992,Makse1995,Makse1998,Pang1995,Makse1996,Ballesteros1999,Ahrens2011,Simon2012,Schrenk2013}
to generate power-law correlated disorder and have yielded estimates for
critical exponents and fractal dimensions characterizing the system in $2D$.
However, they in part used semi-analytical implementations of the FFM,
involving various approximations and free parameters.  In this work we use a
numerical version without free parameters, and whose errors are fully
controlled. 

The remainder of the article is organized as follows:
Section~\ref{secCorrelation} gives a detailed description of the FFM, so that
our implementation is easily reproducible~\footnote{Our code (C++) is available
at \protect\url{github.com/CQT-Leipzig/correlated_disorder}}. Thereafter, in
Sec.~\ref{secMapped}, we specify how the mapping to discrete site variables is
carried out. In the following Sec.~\ref{secPercolation}, we present our results
for the  percolation thresholds on square and cubic lattices. Our main
findings, regarding the fractal dimension $d_{\rm f}$ for long-range correlated
percolation clusters in $2D$ and $3D$, are discussed in Sec.~\ref{secFractal}.
Finally, our results and conclusions are summarized in
Sec.~\ref{secConclusions}.

%
%

\section{Generating long-range correlated disorder}
\label{secCorrelation}

We start with the more general problem of how to obtain a hyper-cubic lattice
$L^D$ of identically distributed random variables $\mys{\vx} \in \mathbb{R}$
that exhibit correlations of the form
\begin{equation}\label{eqCorrelationGoal}
  \avg{ \mys{\vx}\mys{\vx+\vr}}=C_{\vr},
\end{equation}
where $\avg{\ldots}$ denotes the expectation value and $C_\vr$ is a (discrete)
correlation function. $C_{\vr}$ should be symmetric around zero and periodic
along all spatial dimensions, i.e., $C_{\vr+L\vec{e}_i}=C_{\vr}$ for all unit
vectors $\vec{e}_1, \ldots, \vec{e}_D$. It is furthermore convenient to choose
$\mys{\vx}$ as Gaussian random variables with mean $\avg{\mys{\vx}}=\avg{\mys{}}=0$ and
variance $\sigma^2_{\mys{}}=1$. Otherwise, we consider
$C_{\vr}$ to be an arbitrary function for now. (Note that we use the index
notation for explicitly discrete functions).

We use a variant of the Fourier filtering method that employs discrete Fourier
transforms (DFT) and is similar to that from Ref.~\cite{Ahrens2011,Simon2012}.  The key
idea of the FFM is to correlate random variables in Fourier space. The result of
the inverse transform will in general be complex numbers,
$\myphi{\vx}=\myphireal{\vx}+i\myphiimag{\vx}$.  To explain how the method
works, let us now assume that we already have a lattice of complex random
variables $\myphi{\vx}$. Let us further assume that $\{\myphireal{\vx}\}$ and
$\{\myphiimag{\vx}\}$ are independent sets of random variables, each spatially
correlated according to Eq. (3), i.e.,
\begin{align}
  \avg{\myphireal{\vx}\myphireal{\vx+\vr}} &= \avg{\myphiimag{\vx}\myphiimag{\vx+\vr}}=C_{\vr}, \nonumber \\
  \avg{\myphireal{\vx}\myphiimag{\vx+\vr}} &=\avg{\myphiimag{\vx}\myphireal{\vx+\vr}}=0 \label{eqCorrelationPhi},
  \end{align}
and see what that implies for the distributions of Fourier coefficients.

As we are interested in a discrete lattice with periodic boundary conditions of
linear size $L$ and volume $N=L^D$, we consider a DFT of the form
\begin{align}
  \myPhi{\vk} &= \phantom{\frac{1}{N}}\sum_{\vx} \myphi{\vx}\FTxk{\vk\vx}, \\
  \myphi{\vx} &= \frac{1}{N}          \sum_{\vk} \myPhi{\vk}\FTkx{\vk\vx},
\end{align}
where $\sum_{\vx}$ denotes the $D$-dimensional sum over possible realizations of
the vector $\vx$ on the hypercubic lattice. In practice, we employ a numerical fast Fourier
transform (FFT)~\cite{NumericalRecipiesFFT} and follow the convention that
$x_i\in[0,L)$ and $k_i\in[0,L)$.

As shown in Appendix~\ref{secAppCrossCorrelation}, the correlation function is connected to the Fourier coefficients via
\begin{equation}
  2C_{\vr}=
  \avg{\myphi{\vx}^*\myphi{\vx+\vr}^{\phantom{*}}} 
  =\frac{1}{N^2}\sum_\vk\avg{|\myPhi{\vk}|^2}\FTkx{\vk\vr}.
\end{equation}
The discrete spectral density
\begin{equation}
  S_\vk= \sum_\vr C_{\vr}\FTxk{\vk\vr}
\end{equation}
can thus be written as 
%
\begin{align}
  S_\vk 
  &= \sum_\vr \frac{1}{2N^2}\sum_{\vk'}\avg{|\myPhi{\vk'}|^2}\FTkx{\vk'\vr}\FTxk{\vk\vr}\nonumber\\
  &= \frac{1}{2N}\sum_{\vk'}\avg{|\myPhi{\vk'}|^2}\delta_{\vk',\vk}
   = \frac{1}{2N}\avg{|\myPhi{\vk}|^2}\nonumber\\
  &= \frac{1}{2N}\left(\avg{{\myPhireal{\vk}}^2}+\avg{{\myPhiimag{\vk}}^2}\right)
   \label{eqCorrelationCondition}.
\end{align}
In return, this means we can generate {complex} real-space random variables with the
desired correlation from Fourier-space random variables that satisfy
Eq.~\eqref{eqCorrelationCondition}.  It is convenient to consider
distributions of $\myPhi{\vk}$ with zero mean, so that
Eq.~\eqref{eqCorrelationCondition} can be expressed in terms of the variance:
\begin{equation}
 2NS_\vk=\sigma^2_{\myPhi{\vk}}=\sigma^2_{\myPhireal{\vk}}+\sigma^2_{\myPhiimag{\vk}}.
\end{equation}
Hence, we can simply draw real and imaginary parts of $\myPhi{\vk}$
  independently from identical distributions (for each frequency $\vk$):
\begin{equation}
  \myPhiimagreal{\vk}=\sqrt{S_\vk}\myU{},
\end{equation}
where $\myU{}$ is a random variable with mean $\avg{\myU{}}=0$ and variance
$\sigma^2_{\myU{}}=N$. Transforming $\myPhi{\vk}$ back to $\vx$-space, we get
two sets of variables, $\{\myphireal{\vx}\}$ and $\{\myphiimag{\vx}\}$, each
with zero mean and spatial correlations $C_{\vr}$. Thanks to the
orthogonality of the Fourier transform, the two sets are statistically
independent. Each can be associated with the real random site variables
$\mys{\vx}$ in Eq.~\eqref{eqCorrelationGoal} and used for further analysis.
We draw $\myU{}$ from a Gaussian distribution, and so the resulting
distributions will also be Gaussian. (In fact, they would be Gaussian anyway
for large systems due to the central limit theorem.)

The derivation above did not use any assumptions regarding the correlation
function $C_{\vr}$.  However, we see from Eq.~\eqref{eqCorrelationCondition}
that its Fourier transform $S_{\vk}$ needs to be positive. Any $C_{\vr}$ that
is symmetric (around zero) will give rise to real $S_\vk$, but the positivity
constraint is somewhat problematic. For the continuum Fourier transform, it
is in fact also implied by the symmetry~\cite{Makse1996}, but for discrete
systems, some values of $S_\vk$ can become negative. This has to do with the
restricted frequency range, leading to an aliasing effect that causes
periodic modulations on the signal. Note, however, that this is not just an
artifact of the method, but rather implies that some correlations are
fundamentally not possible on a finite discrete lattice.  In practice, we can
simply fix this problem by setting all negative values of $S_{\vk}$ to zero
(``zero-cutoff'').  While this will inevitably modify the resulting
correlations, the effect is usually negligible and vanishes rapidly with
increasing system size, see Appendix~\ref{appSk}.

In short, our version of the FFM can be summarized as follows:
\begin{enumerate}
  \item Choose a discrete correlation function $C_\vr$ that is symmetric around zero.
    For optimal performance, the linear size of the lattice should be $L=2^l$ with integer $l$. 
  \item Perform a DFT, $C_{\vr}\rightarrow S_{\vk}$, and set $S_{\vk}=0$ for all $S_{\vk}<0$ (zero-cutoff). This step only needs to be done once for the whole disorder ensemble.
  \item
    Construct real and imaginary parts of each component independently,
    $\myPhiimagreal{\vk}=\sqrt{S_{\vk}}\myU{}$, where $\myU{}$ is drawn from
    a Gaussian distribution with mean $\avg{U{}}=0$ and variance $\sigma^2_{\myU{}}=N$.

  \item Perform an inverse DFT, $\myPhi{\vk}\rightarrow \myphi{\vx}$,
    to obtain two independent sets of long-range correlated
    variables $\{\myphireal{\vx}\}$ and $\{\myphiimag{\vx}\}$. Each can be
    associated with a set of real random variables $\{\mys{\vx}\}$.
\end{enumerate}
No free parameter is involved in the process. The only minor issue is a
potential zero-cutoff (only for strong correlations), but the practical impact
of this intervention is small and can be assessed a priori (see
Appendix~\ref{appSk}). 

Here we are interested in long-range power-law correlated, Gaussian random variables
with the following properties: 
\begin{align*}
  &\avg{\mys{\vx}\mys{\vx+\vr}}\sim|\vr|^{-a}, \quad\text{for}\quad |\vr|\gg1,\\
  &\sigma^2_{\mys{}}=C_0=1. 
\end{align*}
We follow the suggestion by Makse et al.~\cite{Makse1996} and consider the
correlation function
\begin{equation}\label{eqCorrelationFunction}
  C(\vr) = \left(1+\vr^2\right)^{-a/2},
\end{equation}
which satisfies the above conditions. 
More generally, correlations of the form 
$C(\vr,\alpha)=\left(1+|\vr|^\alpha\right)^{-a/\alpha}$ with $\alpha >0$ are all suitable and
may be chosen depending on the desired behavior of convergence to the asymptotic
limit.

\begin{figure}[t]
  \centering
  \includegraphics[]{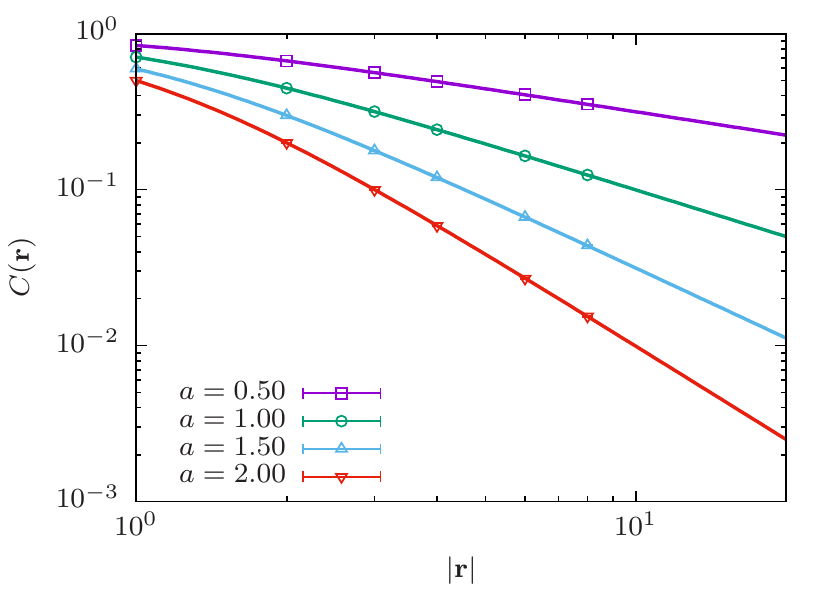}
  \caption{\label{figCorrelationSequence} 
    Correlation function $C(\vr)$ compared to the measured site-site correlation
    function $\avgDis{C_\vr}$ of continuous variables along the $x$-direction on
    a $16^2$ lattice with $R=10^6$ disorder replicas. The continuous random
    variables are obtained via a discrete Fourier transform of $C_\vr$ and
    satisfy $\sigma^2_{\mys{}}=\avgDis{C_0}=1$.
  }
\end{figure}

To verify the correlations numerically, we measure the site-site correlation
function along the ``$x$-direction'' (unit vector $\vec{e}_1$) with periodic
boundary conditions,
\begin{equation}
  \avgDis{C_\vr} =
\avgDis{\frac{1}{N}\sum_{\vx}\left(\mys{\vx}-\avg{\mys{}}
\right)\left(\mys{\vx+|\vr|\vec{e}_1}-\avg{\mys{}} \right)}.
\label{eqCorrelationMeasurement}
\end{equation}
Here $\avgDis{\ldots}$ denotes the disorder average over $R$ replicas, and the
expectation value $\avg{\mys{}}$ is zero, which we verified numerically.
With increasing sample size $R$, the measured correlation function rapidly
converges to the envisaged $C_{\vr}$. As can be seen in
Fig.~\ref{figCorrelationSequence} for a two-dimensional lattice, the agreement
is striking even for very small systems ($16^2$), despite the zero-cutoff.
This is one of the benefits of a fully discrete implementation of the FFM over
semi-analytical techniques, which often cannot faithfully reproduce the desired
distributions for small systems.  For a short review of other variants to
generate long-range power-law correlations and a discussion of some of the
difficulties, see Appendix~\ref{secCorrelationOther}.


\section{Mapping to long-range correlated defects}
\label{secMapped}
\begin{figure}
  \centering
  \includegraphics[]{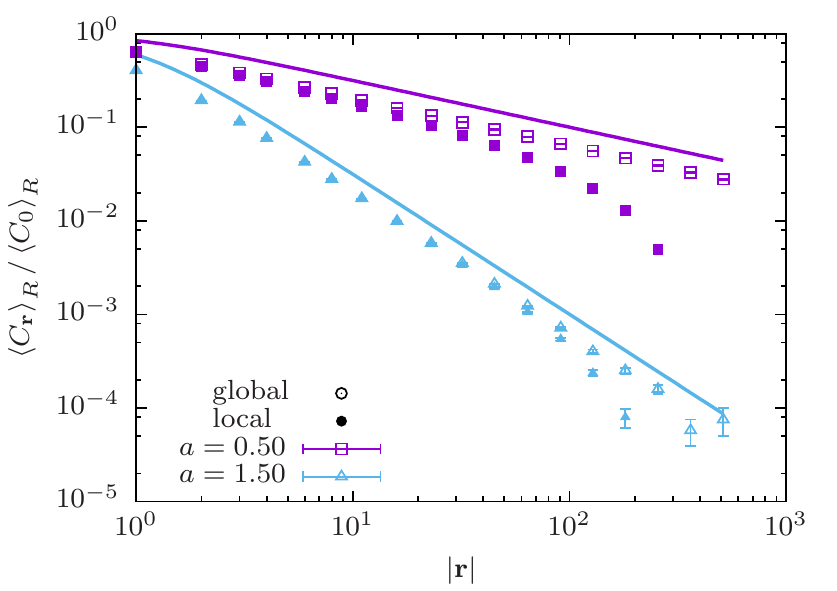}
  \caption{\label{figCorrelationMapped} Normalized correlation function
    of discrete random variables with long-range power-law correlation at the
    percolation threshold $p_c(a)$ (see Table~\ref{tab_pc3D}) on a
    $1024^2$ lattice averaged over $10^4$ disorder replicas. 
    Continuous correlated random variables were obtained as described in
    Sec.~\ref{secCorrelation}. The mapping to discrete variables is performed
    via a \emph{global} approach (open symbols), i.e., on the level of the disorder
    average [see Eq.~\eqref{eqCorrelationMapped}], and via a \emph{local}
    approach, i.e., on the level of each disorder realization (filled symbols).
    The measured site-site correlation function $\avgDis{C_\vr}$ along the
    $x$-axis is normalized with respect to the variance $\avgDis{C_0}$ for
    discrete site variables.
  }
\end{figure}

To study percolation, we have to map the correlated continuous variables
$\mys{\vx}$ to correlated discrete values $\mysd{\vx}\in\{0,1\}$. For this, we
need to specify the mean density of available sites $p$ (considering defects as
$\mysd{\vx}=0$). Here, we use a \emph{global} or
\emph{grand-canonical}~\cite{Wiseman1998} approach and fix the expectation
value $\avg{\sum_\vx\mysd{\vx}/N}=\avg{t}=p$. Therefore, we introduce a threshold
$\theta$ such that sites are considered defects if {$\mys{\vx}<\theta$}. In the
disorder average the $\mys{\vx}$ are {normally} distributed, such that the
threshold is tied to $p$ via
\begin{equation}\label{eqCorrelationMapped}
{p=p(\theta)=\int_{\theta}^{\infty}P(\mys{})d\mys{}=\frac{1}{2}{\rm erfc}\left(\frac{\theta}{\sqrt{2\sigma^2_{\mys{}}}}\right)},
\end{equation}
where erfc denotes the standard complementary error function and
$\sigma^2_{\mys{}}=1$ by construction.  Note that for strong correlations, the
{densities} on individual replica fluctuate significantly. 
%
If we measure the site-site correlation function of discrete variables
according to Eq.~\eqref{eqCorrelationMeasurement} (where we replace
$\avg{\mys{}}$ with $\avg{t}=p$), we observe
$\avgDis{C_0}=\sigma^2_{\mysd{}}<1$. The variance of discrete site variables is
no longer unity but is instead connected to the variance of uncorrelated random
lattices, $\sigma^2_{\mysd{}}=p(1-p)$. 
Figure~\ref{figCorrelationMapped} (open symbols) shows the discrete site-site
correlation function averaged over $10^4$ lattices of size $1024^2$. It can
be seen that the average site-site correlations on discrete lattices mapped via
Eq.~\eqref{eqCorrelationMapped} decay according to
Eq.~\eqref{eqCorrelationFunction} over a long range, though the amplitudes are
somewhat diminished.

Alternatively, one might use a \emph{local} or \emph{canonical}~\cite{Wiseman1998}
approach, adjusting $\sum_\vx\mysd{\vx}/N=p$ for each replica by sorting the
continuous correlated variables and adjusting $\theta$ until \mbox{$\sum_\vx
  \Theta(\mys{\vx} -\theta)/N= p$}, where {$\Theta$} is the unit step
  function~\cite{Schrenk2013}. 
However, fixing $p$ on every lattice tends to suppress correlations on a
macroscopic scale. As can be seen in Fig.~\ref{figCorrelationMapped}~(filled symbols),
this results in a decay rate of the correlation function that is faster than
polynomial.  This effect is most significant for strong correlations and small
systems and can be expected to vanish in the limit of infinite system size. By
contrast, the \emph{global} approach Eq.~\eqref{eqCorrelationMapped} described
above works reliably for any lattice size and appears thus generally
preferable.

\section{Percolation threshold}
\label{secPercolation}
The value of the percolation threshold $p_c$ is not a universal quantity. It may
not only depend on the type of lattice but also on local aspects of the
correlation function $C_\vr$ and hence on the implementation of the FFM.
Numerical results given in this section therefore only apply for the specific
settings we used and cannot be quantitatively compared to those from previous
studies, e.g., Ref.~\cite{Prakash1992}.  We were careful to be explicit about
these settings to ensure that our results for the fractal dimensions are
reproducible, and so that future studies may use our estimates for $p_c$. 

We use the correlation function Eq.~\eqref{eqCorrelationFunction} and perform a
discrete numerical Fourier transform as discussed in the previous section. The
radial distance $|\vr|$ is usually considered in the Euclidean metric, but here
we also use the Manhattan metric, i.e., the minimum number of steps on the
lattice. This is done to demonstrate the sensitivity of $p_c$ to changes of the
correlation function that are not captured in the {correlation parameter} $a$. Later, we also
use the Manhattan metric to test the robustness of our estimates for the
fractal dimensions, which should be the same for both variants.

To define percolation on a finite lattice, we apply the horizontal wrapping
criterion: a cluster percolates if it closes back on itself across one specific
lattice boundary. This choice has the benefit of being translationally
invariant and is known to give relatively small finite-size
errors~\cite{Newman2001}.  The percolation threshold $p_c^L$ for the finite
system of extension $L$ is then defined as the average occupation density at
which a percolating cluster emerges. We estimate this value by determining the
{maximum} threshold $\theta_c$ for each replica of continuous variables at which
a percolating cluster exists for the subset of sites with
{$\mys{\vec{x}}\geq\theta_c$}. We then take the average of the mapped values, 
\begin{equation}\label{eqThresholdMapping}
  p_c^L=\left \langle p\left(\theta_c\right)\right \rangle,
\end{equation}
where the mapping is carried out according to Eq.~\eqref{eqCorrelationMapped}.

\subsection{Square lattice}

\begin{table}
\caption{%
Estimates of the percolation threshold for square lattices with correlated
disorder. For consistency all fits include $L\geq 64$.The extended Harris
criterion, Eq.~\eqref{eqExtendedHarris}, modifies $\nu_a\neq\nu$ for
$a<2/\nu=1.5$.
}\label{tab_pc2D}
\begin{tabular}{ l l l l l l}\hline \hline
  $a$      & $\nu_a$ & $p_c$ (Euclid.)  & $\chi^2/{\rm{dof}}$ &   $p_c$ (Manh.) & $\chi^2/{\rm{dof}}$  \\
  \hline
  $\infty$ & $4/3$~\cite{Nienhuis1984} & $0.592746$~{\cite{Ziff1994}} & & & \\
\hline
  3      & $4/3$ & $0.561406(4)$ & 0.92                        \\
  2.5    & $4/3$ & $0.556214(4)$ & 0.87                        \\
  2      & $4/3$ & $0.550143(5)$ & 0.90 & $0.528397(5)$ & 1.9  \\
  1.75   & $4/3$ & $0.546717(7)$ & 0.41                        \\
  1.5    & $4/3$ & $0.54299(1) $ & 3.5  & $0.519991(8)$ & 4.0  \\
  1.25   & {$8/5$} & $0.53895(2) $ & 1.4                         \\
  1      & $2$   & $0.53452(4) $ & 0.87 & $0.51226(4)$  & 2.28 \\
  0.75   & $8/3$ & $0.5296(1)  $ & 0.63                        \\ 
  0.5    & $4$   & $0.5239(3)  $ & 0.53 & $0.5054(3)$   & 0.66 \\
  0.25   & $8$   & $0.516(1)   $ & 0.38                        \\
  0.1    & $20$  & $0.508(4)   $ & 1.2                         \\
  \hline \hline
\end{tabular}
\end{table}

In $2D$, we extrapolate to the percolation threshold for the infinite system,
$p_c:=p_c^{\infty}$, via the standard finite-size scaling
approach~\cite{Stauffer1992} without higher-order correction terms
\begin{equation}\label{eqFss2D}
{|p_c-p^L_c|}\sim L^{-1/\nu_a}.
\end{equation} 
Here $\nu_a$ denotes the critical exponent of the correlation length. The value
of $\nu_{a}$ is determined by Eq.~\eqref{eqExtendedHarris} with the uncorrelated
correlation-length exponent $\nu=4/3$~\cite{Stauffer1992}.
This assumed behavior of $\nu_a$ has been numerically supported for
percolation on a $2D$ triangular lattice~\cite{Schrenk2013}.

\begin{figure}
  \centering
  \includegraphics[]{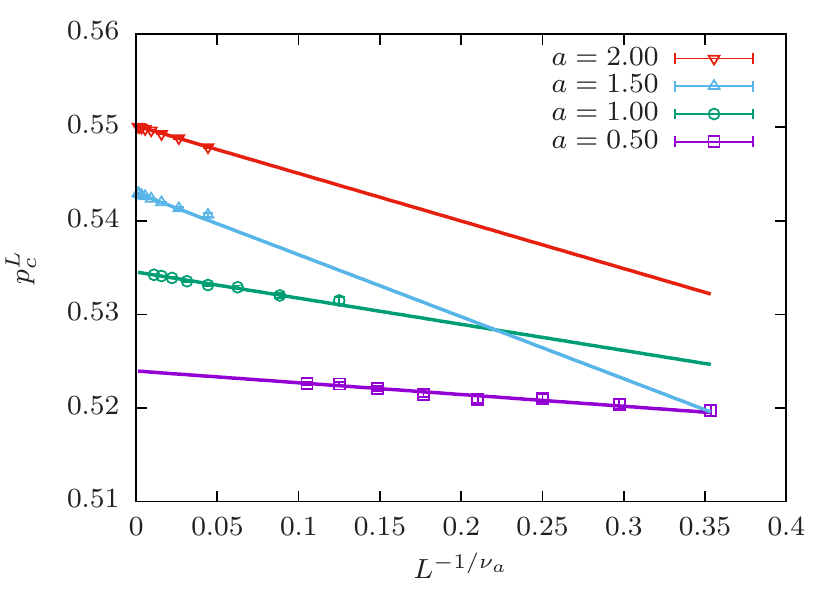}
  \includegraphics[]{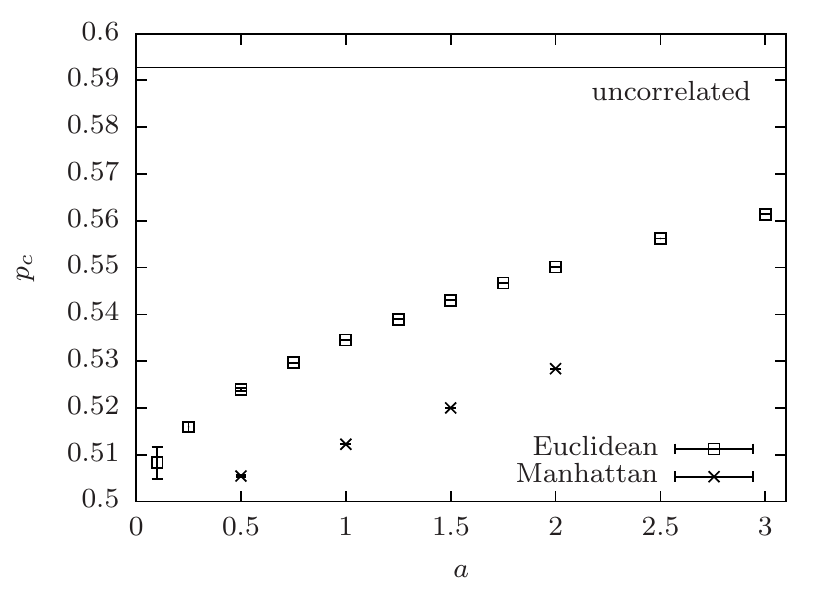}
  \caption{%
    Top: Measured percolation thresholds $p_c^{L}$ for varying lattice size $L$
    and different values of the {correlation parameter} $a$ in $2D$ (Euclidean metric).  Lines
    show the best fits of Eq.~\eqref{eqFss2D} to the data, whose intercepts
    represent our estimates for the infinite-system value $p_c$. Bottom:
    Estimates for $p_c$ as function of the correlation strength for the square
    lattice with distances measured in the Euclidean metric (squares) and the
    Manhattan metric (crosses). The horizontal line shows the value for the
    system without correlations.
  }\label{figPerc2DFss}
  
\end{figure}

To obtain our numerical estimates, we randomly generated $10^5$ replicas for
each size $L=2^l$ where $l=6,\ldots,13$ ($L=64$--$8192$). Some of the results for $p^L_c$ are shown in
Fig.~\ref{figPerc2DFss} (top) together with least-squares fits of
Eq.~\eqref{eqFss2D} over the range $L\geq64$. The estimates for $p_c$ are the
$y$-intercepts of the fit curves.  The values are listed in
Table~\ref{tab_pc2D}, where we also give the reduced $\chi^2$-values per
degree-of-freedom (dof) of the respective fits. The last columns show our
results for systems where the Manhattan metric is used to set the distance
$|\vec{r}|$ for the correlations.  Here the estimates for $p_c$ are considerably
smaller than for the Euclidean case, underlining the strong dependence of $p_c$
on the details of the correlation function. In both cases the $\chi^2$-values
are mostly close to one, justifying the simple scaling ansatz. However, they are
quite large at the ``crossover'' value of $a_{\rm H}=1.5$, where the behavior is supposed
to change according to the extended Harris criterion,
Eq.~\eqref{eqExtendedHarris}. {This suggests the presence of additional correction terms in the vicinity of $a_{\rm H}$, possibly of logarithmic nature.} 

An overview of the results for the percolation thresholds as a function of $a$
is shown in the bottom plot of Fig.~\ref{figPerc2DFss}. As can be seen,
correlations tend to lower $p_c$, which is intuitive as they promote the
emergence of larger clusters. As noted in Ref.~\cite{Prakash1992} the value of
$p_c$ for the square lattice must eventually approach $1/2$. This bound can be
understood considering that a cluster of occupied sites that wraps the system in
one direction exists if and only if no cluster of defects wraps the system in
the orthogonal direction, where the defects are allowed to connect via
next-nearest neighbors (diagonally). For $a\longrightarrow 0$, the relevance of
these next-nearest neighbor connections becomes negligible, and the resulting
symmetry between clusters of defects and occupied sites demands $p_c=1/2$. Note
that for the Manhattan metric, diagonal correlations are weaker to begin with.
The strong deviations do not only depend on the chosen metric but are already
affected by the details of the employed method, as can be seen by comparing to
results we obtained with the continuous FFM on a square
lattice~\cite{Fricke2017}, which qualitatively look similar but do not agree
within error bars.

When $a$ is increased, i.e., when the correlation strength is diminished, $p_c$
must converge towards the value for the uncorrelated system as long as
$C(\vr)/C(0)\longrightarrow 0$ for all $|\vr|>0$.
Note, however, that the uncorrelated value is only reached in the limit $a\rightarrow \infty$ and not
at $a=D$, where the correlations become effectively short range.
This is contrary to the results from previous studies due to differing
definitions of the correlation function $C(\vr)$, which at $a=2$ has a vanishing
amplitude in Ref.~\cite{Prakash1992} and a divergent variance $C(0)$ in
Ref.~\cite{Pang1995}.

\subsection{Cubic lattice}
\label{secPercolation3D}
The version of the FFM described in Sect.~\ref{secCorrelation} can directly be
applied in three (or more) dimensions as well, which allowed us to study
percolation with long-range correlated disorder on the cubic lattice. We
looked at systems with linear extensions in the range $L=8$--$512$, and we
again generated $10^5$ random replicas for each size. Unlike in $2D$,
however, the simple finite-size scaling approach {to estimate the percolation threshold $p_c$}, Eq.~\eqref{eqFss2D}, proved
unsuccessful, suggesting the need of higher-order terms (see Ref.~\cite{Herrmann1983}
for a discussion of finite-size scaling for uncorrelated systems):
\begin{equation}
{|p_c-p^L_c|}\sim L^{-1/\nu_a}(A+B L^{-\omega}+C L^{-1/\nu_a}+\ldots),
\end{equation}
where $\nu_a$ is given by Eq.~\eqref{eqExtendedHarris} with $\nu$ the
correlation-length exponent for uncorrelated percolation
(0.8764(12)~\cite{Wang2013}, 0.8762(12)~\cite{Xu2014},
0.8751(11)~\cite{Hu2014}).

\begin{table}[]
\caption{%
  Estimates of the percolation threshold for cubic lattices with correlated
  disorder. For consistency all fits include $L\geq 32$. The extended Harris
  criterion, Eq.~\eqref{eqExtendedHarris}, modifies the {correlation-length exponent} $\nu_a\neq\nu$ for $a<2/\nu
  \approx 2.28$.
}\label{tab_pc3D}
\begin{tabular}{ l l l l l l}\hline \hline
  $a$      & $\nu_a$ & $p_c$ (Euclid.)  & $\chi^2/{\rm{dof}}$ &   $p_c$ (Manh.) & $\chi^2/{\rm{dof}}$  \\
  \hline
  $\infty$ & $0.8762$ & 0.311610(2) & 0.44 & & \\
  \hline
  4        & $0.8762$ & 0.238778(4) & 0.10 & & \\
  3        & $0.8762$ & 0.208438(5) & 0.83 & 0.209315(4) & 0.75 \\
  2.5      & $0.8762$ & 0.188289(7) & 1.9  & 0.189801(5) & 6.0  \\
  2        & $1$    & 0.16302(2)  & 3.0   & 0.16514(1)  & 0.54 \\
  1.5      & $4/3$  & 0.13022(5)  & 0.51  & 0.13251(4)  & 0.37 \\
  1        & $2$    & 0.0863(3)   & 1.1   & 0.0878(2)   & 0.86 \\
  0.5      & $4$    & 0.025(3)    & 1.4   & 0.030(2)    & 0.25 \\
  \hline \hline
\end{tabular}
\end{table}

\begin{figure}
  \centering
  \includegraphics[]{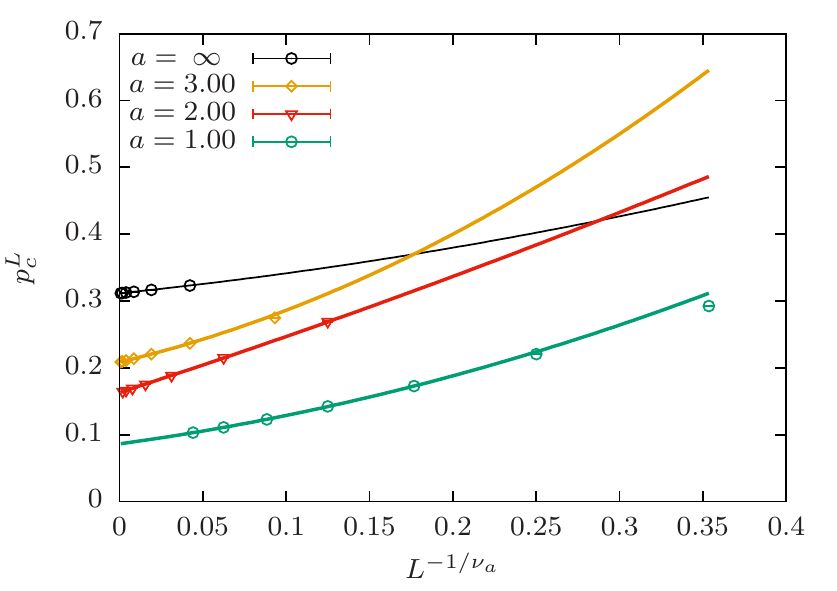}
  \includegraphics[]{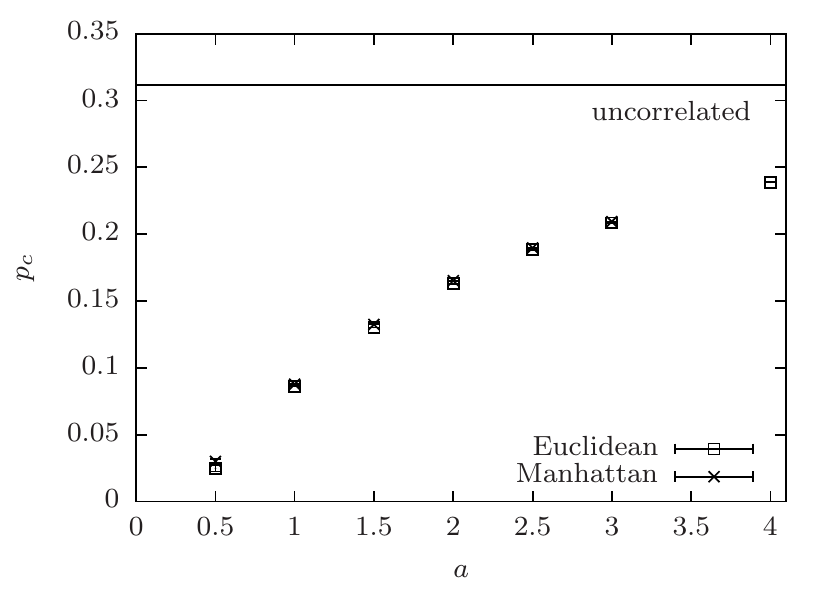}
  \caption{%
    Top: Measured percolation thresholds $p_c^{L}$ for varying lattice size $L$
    and different values of the {correlation parameter} $a$ in $3D$ (Euclidean metric). Lines
    show the best fits of Eq.~\eqref{eqFss3D} to the data, whose intercepts
    represent our estimates for the infinite-system value $p_c$. Bottom:
    Estimates for $p_c$ as function of the correlation strength for the cubic
    lattice with distances measured in the Euclidean metric (squares) and the
    Manhattan metric (crosses). The horizontal line shows the value for the
    system without correlations.
  }\label{figPerc3DFss}
  
\end{figure}

In practice, the correction to Eq.~\eqref{eqFss2D} seems to be described well by
the latter (quadratic) term alone, suggesting that the correction-to-scaling
exponent $\omega$ is relatively large. This is in fact the case for the
uncorrelated system, where previous estimates locate the correction-to-scaling
exponent between $\omega\approx 1.62$~\cite{Ballesteros1999} and $\omega\approx
1.2$~\cite{Wang2013}. 
We thus used the ansatz
\begin{equation}\label{eqFss3D}
{|p_c-p^L_c|}\sim L^{-1/\nu_a}(A+C L^{-1/\nu_a}),
\end{equation}
which we fitted to the data for $L\geq32$.
The corresponding fit curves and our results for $p_c^L(a)$ are shown for
selected correlations in Fig.~\ref{figPerc3DFss} (top), and the resulting
estimates for $p_c=p_c^{\infty}$ are listed in Table~\ref{tab_pc3D}. Again, we
see that the changing behavior predicted by the extended Harris criterion (at
$a_{\rm H}=2/\nu\approx2.28$) manifests itself in a poorer quality of the fits {for nearby values ($a=2$ and $a=2.5$)}.
Our estimate for the uncorrelated case, $p_c(\infty)=0.311\,610(2)$, is in decent
agreement with previous estimates
($0.311\,608\,1(11)$~\cite{Ballesteros1999}, $0.311\,607\,7(2)$~\cite{Wang2013}, 0.311 607 68(15)~\cite{Xu2014}). 

In contrast to the $2D$ situation, using the Manhattan metric in place of the Euclidean
metric to measure the distance for the correlation function does not
significantly lower the percolation threshold. As can be seen in
Table~\ref{tab_pc3D} and Fig.~\ref{figPerc3DFss} (bottom), the values are even
slightly larger. That is plausible since the argument why the Manhattan metric
should lower $p_c$ in $2D$ does not apply in $3D${, where wrapping clusters of defects and occupied sites can coexist}.
This also means
that there is no obvious lower bound for $p_c$ in $3D$ other than zero. Indeed,
our estimates for strong correlations are very small, and the overview shown in
Fig.~\ref{figPerc3DFss} (bottom) even seems to suggest the extrapolation
$p_c\rightarrow0$ for $a\rightarrow 0$. 

We should note, however, that the scaling ansatz Eq.~\eqref{eqFss3D} is mainly
motivated empirically. Especially for small $a$, some of the finite-size
corrections have a different origin as in the uncorrelated system, namely that
smaller systems are not self-averaging: For small $a$ and small $L$, the
continuous {site} variables $\mys{\vx}$ within each
{individual} replica tend to be very similar, and about half the
ensemble has mostly negative values, while the other has mostly positive values.
In the limit $a\rightarrow 0$ (at fixed $L$) the values $\mys{\vx}$ across each
replica {become} constant, so that a wrapping cluster
{in the discrete system emerges when a threshold
$\theta_c=\mys{\vx}$ is used for the mapping}. Since
the overall distribution of the $\mys{}$-values is symmetric (Gaussian) and we
define $p^L_c$ according to Eq.~\eqref{eqCorrelationMapped} and
Eq.~\eqref{eqThresholdMapping}, the $a\rightarrow 0$ limit at fixed $L$ is
$p^L_c(0)=1/2$. This ``segregation'' finite-size effect might play a significant
role for the most strongly correlated cases ($a=\{0.5,1\}$), and our respective
estimates should therefore be taken with a pinch of salt.




\begin{table}[]
\caption{%
  Fractal dimension for square lattices with correlated disorder. For
  consistency all fits include sizes $L\geq128$. In two dimensions we consider
  only the leading-order behavior at the finite-size percolation transition.
  Results from Euclidean and Manhattan metric are in good agreement.
}\label{tab_frac2D}
\begin{tabular}{ l l l l l}\hline \hline
  $a$ & $d_{\mathrm{f}}$ (Euclid.)  & $\chi^2/{\rm{dof}}$  & $d_{\mathrm{f}}$ (Manh.)  & $\chi^2/{\rm{dof}}$\\
  \hline
  $\infty$ & \multicolumn{2}{l}{$91/48\approx1.89583...$}\\
  \hline
  3      &  1.8961(2)  & 0.74   &           &      \\
  2.5    & 	1.8962(2)  & 1.2    &           &      \\
  2      &  1.8966(2)  & 4.5    & 1.8964(2) & 1.2  \\
  1.75   &  1.8964(2)  & 2.8    &           &      \\
  1.5    &  1.8965(3)  & 1.6    & 1.8956(3) & 2.3  \\
  1.25   &  1.8950(3)  & 1.2    &           &      \\
  1      &  1.8961(3)  & 1.2    & 1.8952(3) & 0.29 \\ 
  0.75   &  1.9006(4)  & 1.2    &           &      \\ 
  0.5    &  1.9128(5)  & 0.47   & 1.9126(6) & 0.17 \\
  0.25   &  1.9360(6)  & 0.085  &           &      \\
  0.1    &  1.9602(8)  & 0.39   &           &      \\
   \hline \hline
\end{tabular}
\end{table}

\section{Fractal Dimension}\label{secFractal}
The fractal dimension $d_{\mathrm{f}}$ describes how the volume of a critical
percolation cluster increases with its linear size. It can conveniently be
estimated via
\begin{equation}\label{eqDf}
\left \langle V \right \rangle \sim L^{d_{\mathrm{f}}},
\end{equation}
where $L$ is the lattice extension and $\left \langle V \right \rangle$ denotes
the average number of sites in the largest cluster~\cite{Stauffer1992}. Note
that for correlated systems, it is important to include replicas with no
percolating cluster.  It is possible to either consider all systems at the same,
asymptotic concentration $p_c$ or to take size-dependent values, $p^L_c$. We
opted for the latter approach, so we would not have to rely on the fitting
ansatz for $p_c$.

\begin{figure}
  \centering
  \includegraphics[]{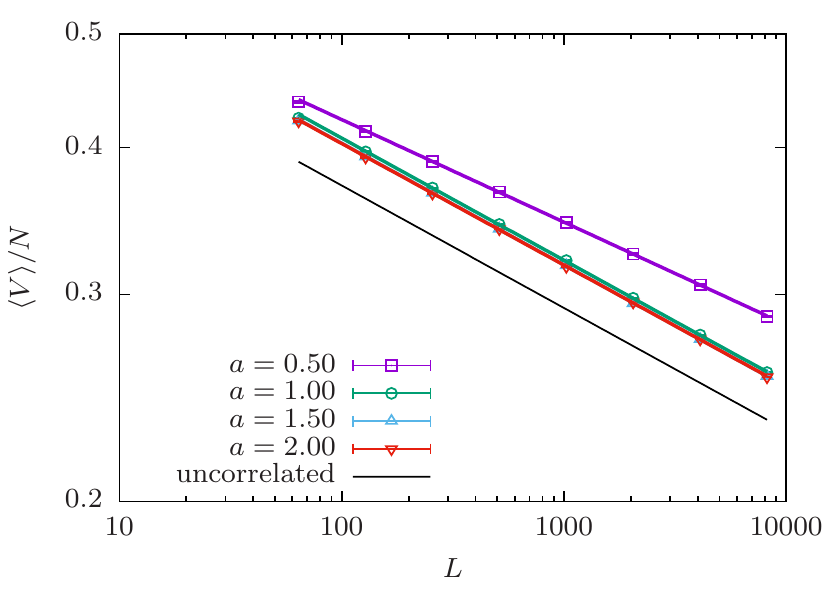}
  \includegraphics[]{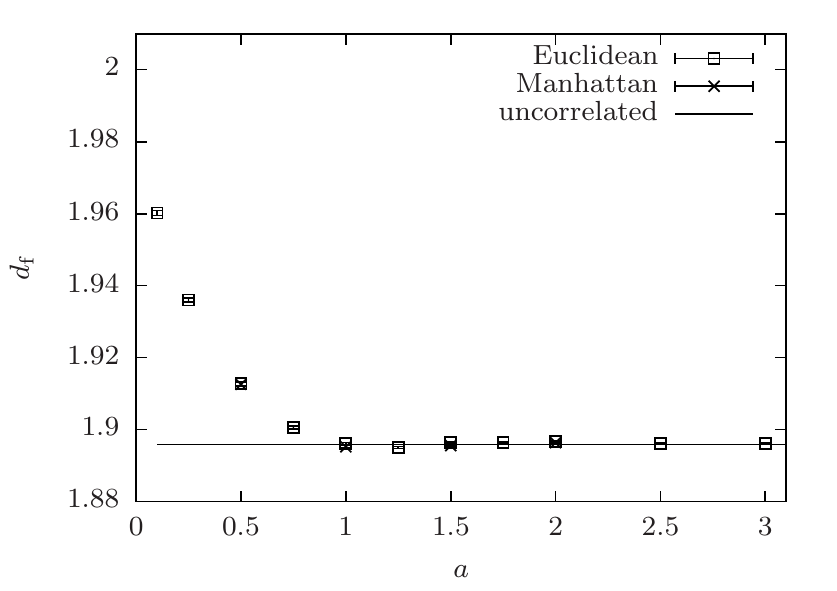}
  \caption{Top: Average volume fraction of the largest cluster vs. lattice
  extension in $2D$ plotted for different correlations $a$ on a
  double-logarithmic scale (Euclidean metric). Colored lines are least-squares
  fits to Eq.~\eqref{eqDf}.  The black line represents the behavior of the
  uncorrelated system with a slope of $91/48-2$. Bottom: Overview of our
  estimates for the fractal dimensions as function of $a$ with the horizontal line again corresponding to the uncorrelated
  system.}\label{figFrac2DFss}
\end{figure}  

\subsection{Square lattice}
In $2D$, finite-size corrections again turned out to be small, so that fitting
Eq.~\eqref{eqDf} without any higher-order correction terms worked well.
Figure~\ref{figFrac2DFss} (top) shows the average volume of the largest cluster
relative to the total number of sites, $\left \langle V \right \rangle/L^D$,
for several different values of $a$ plotted on a double-logarithmic scale. The
lines correspond to least-squares fits of Eq.~\eqref{eqDf} over the range
$L\geq 128$, and their slopes show the differences to the Euclidean dimension,
$d_{\mathrm{f}}-2$. Our resulting estimates for $d_{\mathrm{f}}$ can be found
in Table~\ref{tab_frac2D} together with the reduced $\chi^2$-values of the
fits.  Also listed are estimates obtained using the Manhattan instead of the
Euclidean metric. Here, the fits yielded smaller amplitudes, but the exponents
resulted very similar. This can be seen in Fig.~\ref{figFrac2DFss} (bottom),
which shows an overview of the estimates for $d_{\mathrm{f}}$.  The data
verifies that $d_{\rm f}$ is universal, i.e., independent of system details.
For weak correlations the uncorrelated value, $d_{\rm
f}=91/48$~\cite{Nienhuis1984}, seems to be recovered in accordance with the
extended Harris criterion [Eq.~\eqref{eqExtendedHarris}] and earlier numerical
findings~\cite{Prakash1992,Makse1995a,Makse1998,Schrenk2013}.
Interestingly though, there seems to be no increase of $d_{\mathrm{f}}$ directly
below $a_{\rm H}=3/2$, the crossover threshold set by the extended Harris
criterion. {For the Manhattan
metric, the fit quality is still diminished around $a_{\rm H}$, suggesting that
the threshold may still affect correction terms. However, it is yet unclear why $\chi^2$ is
largest at $a=2$ for the Euclidean case.} 

{The fact that $d_{\mathrm{f}}$ does not increase directly
below $a_{\rm H}$ was already noted} in Ref.~\cite{Schrenk2013}, where a
crossover threshold of $a_{\rm x}=2/3$ (or in terms of the Hurst exponent
$H_{\rm x}=-a_{\rm x}/2=-1/3$) was suggested instead. However, that value is
not quite consistent with our findings, which show a significant increase
already at $a>3/4$. Another disagreement regards the behavior in the correlated
limit, $a\rightarrow 0$ ($H\rightarrow 0$): our results are consistent with the
idea that the fractal dimension converges to the ``Euclidean'' value of $D=2$
as clusters get more and more compact, while
according to Ref.~\cite{Schrenk2013} the value stays well below $2$. This
discrepancy may be owed to the use of different mapping rules as discussed at
the end of Sec.~\ref{secMapped}. 

{It is interesting to compare the results for $d_{\rm f}$ with the Ising model at criticality,
which exhibits spin-spin correlations of the form $\langle S_i S_j\rangle\sim
r^{-(d-2+\eta)}$. In two dimensions $\eta=1/4$ and the fractal dimension
of the geometrical Ising clusters is  $d_\mathrm{f}=187/96=1.9479\ldots$~\cite{Stella1989,Duplantier1989}, which is indeed quite similar
to our result of $d_\mathrm{f}=1.9360(6)$ for $a=1/4$. As already noted~\cite{Prakash1992},
the results could not be expected to agree perfectly. In fact, it is intuitive that $d_\mathrm{f}$ should be slightly larger for Ising clusters, where the spin-spin correlation function is 
essentially the probability that two spins belong to the same cluster. In our system, by contrast, spins from unconnected clusters still contribute to the correlation function, so that connected clusters may be ``thinner'' for the same decay exponent.  
}

\begin{table}[]
\caption{%
  Fractal dimension for cubic lattices with correlated disorder. For consistency
  all fits include sizes $L\geq16$. In three dimensions, we require higher-order
  corrections of the form Eq.~\eqref{eqDfFSS}. Assuming universality, we
  perform simultaneous fits including both Euclidean and Manhattan metric.
}\label{tab_frac3D}
\begin{tabular}{ l l l l}\hline \hline
  $a$      & $d_{\mathrm{f}}$ & $w $ & $\chi^2/{\rm{dof}}$  \\
  \hline
  $\infty$   & 2.52295(15)~\cite{Wang2013} & 1.2(2)~\cite{Wang2013} &  \\
  \hline
  3        & 2.524(2) & 1.2(2)  & 0.70 \\
  2.5      & 2.522(2) & 0.9(2)  & 3.3  \\
  2        & 2.512(3) & 0.9(2)  & 1.9  \\
  1.5      & 2.507(6) & 0.6(1)  & 0.55 \\
  1        & 2.6(2)   & 0.20(1) & 2.1  \\
  \hline \hline
\end{tabular}
\end{table}

\begin{figure}
  \centering
  \includegraphics[]{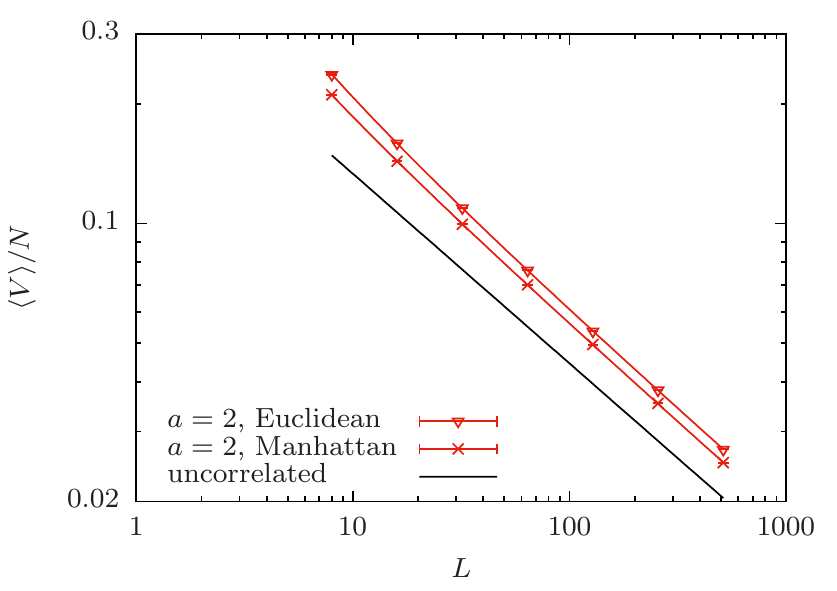}
  \includegraphics[]{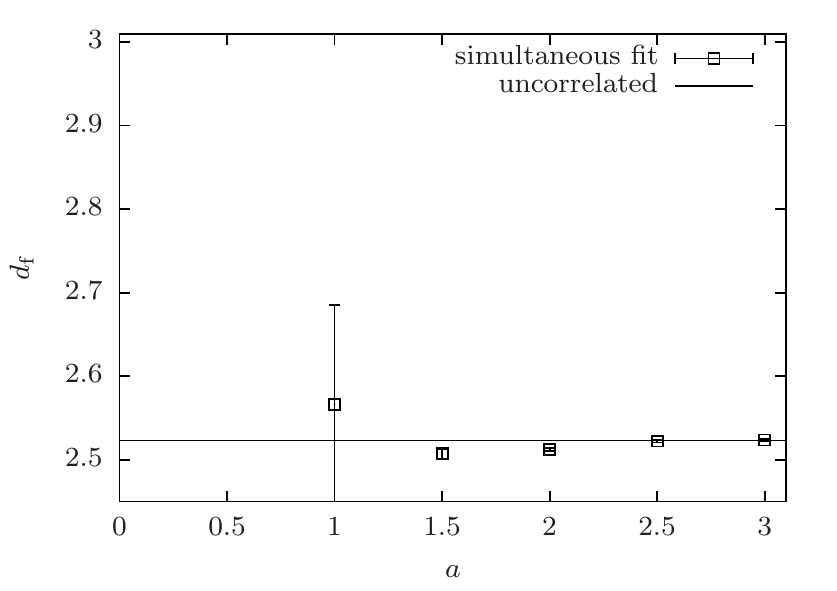}
  \caption{Top: Average volume fraction of the largest cluster vs. lattice
    extension in $3D$ plotted for the exemplary case $a=2$ on a
    double-logarithmic scale. Colored lines are the simultaneous fits according
    to Eq.~\eqref{eqDfFSS}; the black line represents the behavior of the
    uncorrelated system with a slope of $2.52295-3$. Bottom: Overview of our
    estimates for the fractal dimensions as function of $a$ with the horizontal
    line again corresponding to the uncorrelated system. Our fitting did not
    work properly for very strong correlations ($a\leq 1$).
}\label{figFrac3DFss}
  
\end{figure}  

\subsection{Cubic lattice}
The situation in $3D$ turned out to be more difficult. As with the percolation
threshold, the scaling behavior seems to involve strong finite-size corrections,
so that simply fitting Eq.~\eqref{eqDf} would not work for the system sizes that
we considered. Including a correction term also failed as the
fit could not handle two additional parameters. What did work reasonably
well, at least for $a>1$, was fitting our data for the Euclidean and the
Manhattan versions simultaneously, while assuming the exponents of the leading
term and the correction to be equal for both cases:
\begin{align}
\left \langle V \right \rangle_{\mathrm{Euclid.}} & = A_1L^{d_{\mathrm{f}}}(1+B_1L^{-w}),\nonumber\\
\left \langle V \right \rangle_{\mathrm{Manh.}}  & = A_2L^{d_{\mathrm{f}}}(1+B_2L^{-w})\label{eqDfFSS}.
\end{align}
{This approach was motivated by general universality arguments~\cite{Cardy1996, Zinn-Justin2007} and our previous observation that the fractal
dimensions in $2D$ are the same for both versions. We assume that equality also
holds for the correction exponents $\omega$, which seems reasonable since
$\omega$ is also strongly believed to be universal for percolation without correlations, see for instance Ref.~\cite{Ballesteros1999a}.}
Figure~\ref{figFrac3DFss}
shows finite-size scaling data and fits for the case $a=2$ as an example (top)
and an overview of the obtained estimates for $d_{\mathrm{f}}$ (bottom). The
values of our estimates can be found in Table~\ref{tab_frac3D} together with the
correction exponents and the $\chi^2$-values of the fits.  Unfortunately, the
data for $a\leq 1$ could not be convincingly fitted by this approach. For these
strongly correlated cases, one would probably need to investigate systems still
much larger than $512^3$. For $a\geq 1.5$, the value for $d_{\mathrm{f}}$ seems
to be very similar to the one without correlations ($d_{\rm
f}(\infty)=2.522\,95(15)$~\cite{Wang2013}).  As in $2D$, the Harris threshold,
$a_{\rm H}=2/\nu\approx2.28$, does hence not determine the onset of a sudden increase
in the fractal dimension.  Surprisingly, the value even seems to decrease
slightly below $a_{\rm H}$.  At close inspection, this can also be observed in $2D$
for $a=1.25<a_{\rm H}$, compare Table~\ref{tab_frac2D}.  A diminishing fractal
dimension does not seem plausible as stronger correlations should make the
clusters more compact. We suspect that a correction term comes into play at
$a_{\rm H}$ which is not captured by our fitting approach.

\section{Conclusions}~\label{secConclusions}
We presented high-precision results for the percolation thresholds on square
and cubic lattices with long-range power-law correlated disorder as well as
estimates for the fractal dimensions of the critical percolation clusters. The
correlations were generated using the Fourier filtering method (FFM) based on
the discrete Fourier transform. We specified the details of our implementation,
so that it may easily be reproduced~\cite{Note1} and discussed the differences to previous
approaches regarding, e.g., how the continuous site variables are mapped to
discrete disorder. 

The percolation threshold is dependent on the employed method and moreover on
short-range details of the model. We demonstrated this by using both the
standard Euclidean metric and the discrete Manhattan metric to define the
correlation. 
The effect of this choice on the percolation threshold is particularly strong
for the square lattice.  This is because diagonal correlations are weaker for
the Manhattan metric, bringing the system closer to $p_c=0.5$ where the
percolation thresholds for occupied sites and defects connected via
next-nearest (diagonal) neighbors coincide.  In general, correlations were
shown to lower $p_c$, and in three dimensions the value even becomes very close
to zero for small $a$, i.e., strong correlations.

The fractal dimension, by contrast, is a universal quantity and does not depend
on details of the model. We verified that for large $a$ ({weak} correlations) the
fractal dimension of the uncorrelated model is reproduced showing $d_{\rm
f}\approx91/48$~($2D$) and $d_{\rm f}\approx 2.52$~($3D$). This was expected above
the bound from the extended Harris criterion, i.e., for all $a\geq a_{\rm H}$ with
$a_{\rm H}=1.5$~($2D$) and $a_{\rm H}\approx2.28$~($3D$). However, as was previously noticed for
the triangular lattice~\cite{Schrenk2013},  $d_{\rm f}\approx d_{\rm f}^{\rm
uncorr}$ seems to remain true also well below the Harris bound. In two
dimensions, our data suggests that the value of $d_{\rm f}$ starts to rise below
$a_{\rm x}\approx 1$, approaching $d_{\rm f}\rightarrow 2$ as $a\rightarrow 0$.
Differences to previous findings may be attributed to different mapping
prescriptions employed.  To obtain estimates for $d_{\rm f}$ in three
dimensions, we simultaneously fitted our data for correlations with Euclidean
and Manhattan metric using a polynomial fit with a correction term.
Unfortunately, this approach did not work for very strong correlations, i.e.,
for $a\leq 1$.  In the accessible range, the values were found very close to the
uncorrelated value. Below $a_{\rm H}\approx2.28$, they even resulted slightly smaller,
which we suspect is due to changing corrections to scaling. We conclude that
while the bound from the Harris criterion does not seem to determine a change in
the leading exponent $d_{\rm f}$, it does affect the system's sub-leading
behavior.

\begin{acknowledgments}
We thank Martin Weigel and Martin Treffkorn for helpful discussions.  This work
has been supported by an Institute Partnership Grant “Leipzig-Lviv” of the
Alexander von Humboldt Foundation (AvH). 
  Further financial support from the Deutsche Forschungsgemeinschaft (DFG) via the Sonderforschungs\-bereich SFB/TRR
102 (project B04), the Leipzig Graduate School of Natural Sciences
``BuildMoNa'', as well as from the Deutsch-Franz\"osische Hochschule (DFH-UFA)
through the Doctoral College ``${\mathbb L}^4$'' under Grant No.\ CDFA-02-07
and the EU through the Marie Curie IRSES network DIONICOS under Contract No.\
PIRSES-GA- 2013-612707 (FP7-PEOPLE-2013-IRSES) is gratefully acknowledged.
J.~Z.\ received financial support from the German Ministry of Education and
Research (BMBF) via the Bernstein Center for Computational Neuroscience (BCCN)
G{\"o}ttingen under Grant No.~01GQ1005B.
\end{acknowledgments}

\begin{appendix}
\section{Discrete Wiener-Khinchin theorem}
\label{secAppCrossCorrelation} 
We require $\myphi{\vx}$ to be complex random variables with independent real
and imaginary contributions. For a given disorder realization the
  lattice average of $\myphi{\vx}^*\myphi{\vx+\vr}^{\phantom{\ast}}$  can be written as 
\begin{align}
  \frac{1}{N}&\sum_\vx \myphi{\vx}^*\myphi{\vx+\vr}^{\phantom{*}} \nonumber \\ 
    &= \frac{1}{N}\sum_\vx \left( \frac{1}{N}\sum_\vk \myPhi{\vk}^*\FTxk{\vk\vx}\right) \left(
    \frac{1}{N}\sum_\vl \myPhi{\vl}\FTkx{\vl(\vx+\vr)}\right) \nonumber \\
    &=\frac{1}{N^2}\sum_\vk\myPhi{\vk}^*\sum_\vl\myPhi{\vl}\FTkx{\vl\vr}\frac{1}{N}\sum_\vx\FTxk{(\vk-\vl)\vx} \nonumber \\
    &=\frac{1}{N^2}\sum_\vk\myPhi{\vk}^*\sum_\vl\myPhi{\vl}\FTkx{\vl\vr}\delta_{\vl,\vk} \nonumber \\
    &=\frac{1}{N^2}\sum_\vk|\myPhi{\vk}|^2\FTkx{\vk\vr}. \label{WKT}
\end{align}
Here, we used the notation of a $D$-dimensional Kronecker-Delta function
  \mbox{$\delta_{\vec{l},\vec{k}}=\Pi_i\delta_{l_i,k_i}=\frac{1}{L^D}\sum_\vx
  e^{2\pi i(\vec{k}-\vec{l})\vx/L}$}.
The result is essentially the discrete Wiener-Khinchin theorem, a special case of the
cross-correlation theorem.
%
%

Taking the disorder average on both sides of Eq.~\eqref{WKT} and exploiting translational invariance on the left, we thus obtain
\begin{align}
  \avg{\myphi{\vx}^*\myphi{\vx+\vr}^{\phantom{*}}} 
  &=\frac{1}{N^2}\sum_\vk\avg{|\myPhi{\vk}|^2}\FTkx{\vk\vr}.
\end{align}

\section{Effect of zero-cutoff in $S_\vk$ on $C_\vr$}
\label{appSk}
As mentioned in Sec.~\ref{secCorrelation}, particular choices of $C(\vr)$
evaluated on a finite lattice may lead to unphysical negative values of the
discrete spectral density $S_\vk$. This seems to occur only for strong
correlations (small $a$) and becomes more noticeable with increasing
dimension.  Numerically, we deal with this issue by a zero-cutoff, i.e., by using a modified spectral density
\begin{equation}
  \widetilde{S}_\vk= \begin{cases}
  S_\vk, \quad & \mathrm{if}\:S_\vk \geq 0\\  
  0,          & \mathrm{else.}
  \end{cases}    
\end{equation}
This inevitably affects the
resulting correlation function.
We can directly predict the effect from the inverse discrete Fourier transform
of $\widetilde{S}_\vk$ since 
\begin{equation}
  \widetilde{C}_\vr= \frac{1}{N}\sum_{\vk} \widetilde{S}_{\vk}\FTkx{\vk\vr}
  \label{eqCorrelationDeviation}
\end{equation}
is of course the asymptotic limit of the measured site-site correlation
function $\avgDis{C_\vr}$ for large sample size $R$.  

\begin{figure}
  \includegraphics{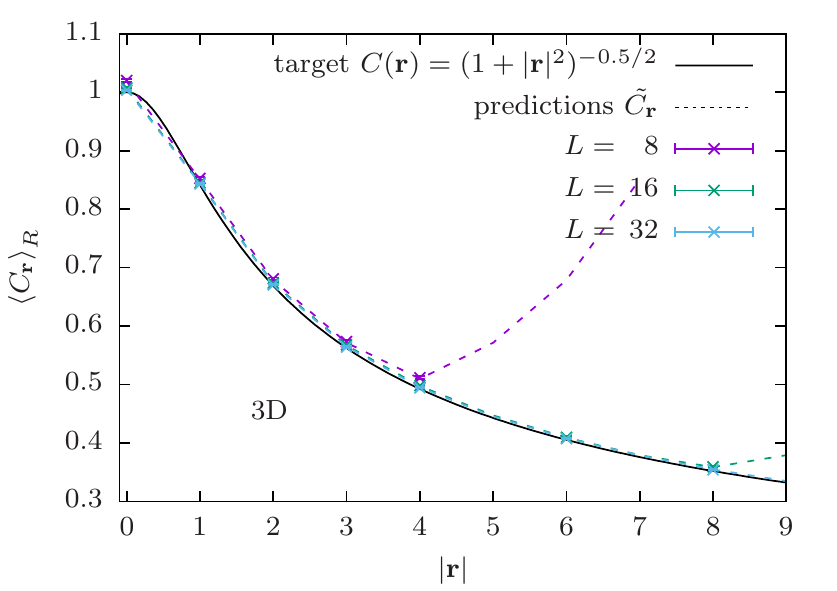}
  \includegraphics{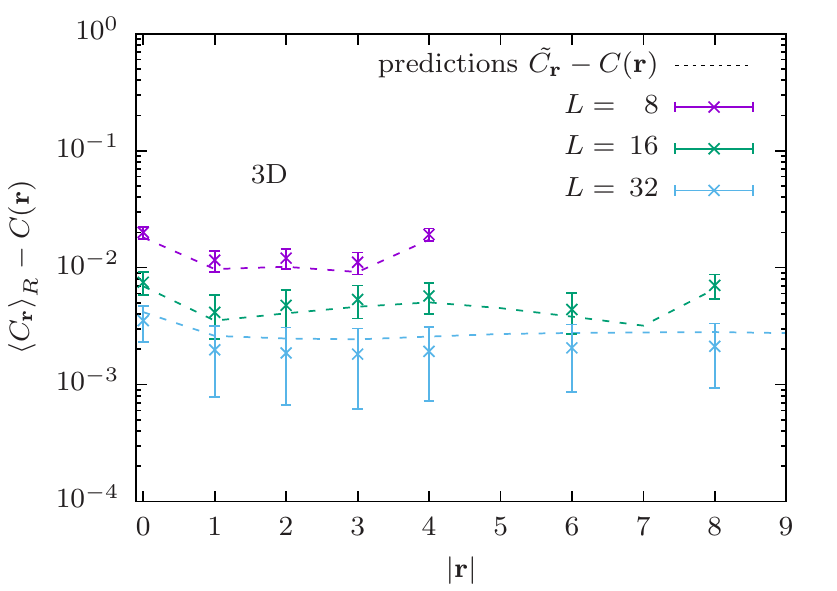}
  \caption{%
    (Top) Demonstration of the effect of the zero-cutoff
    $S_\vk\rightarrow\widetilde{S}_\vk$ on the measured correlation function
    $\avgDis{C_\vr}$ in three dimensions for $a=0.5$. Also shown is the
    prediction $\widetilde{C}_\vr$ for the asymptotic limit from
    Eq.~\eqref{eqCorrelationDeviation}.
    (Bottom) Plot of the deviations from the desired function $C(\vr)$. 
    \label{figCorrelationZeroCutoff}
  }
\end{figure}

It turns out that the effect of this zero-cutoff is very small, if present at
all, with deviations mainly occurring for small lattices. With increasing
lattice size, the predicted (and measured) deviations quickly converge towards
the desired correlation function. To demonstrate this, we consider the example
of three-dimensional lattices with (strong) correlations $a=0.5$ and linear
extensions $L=\{8,16,32\}$ in Fig.~\ref{figCorrelationZeroCutoff}. The measured
correlation function of continuous site variables along the $x$-direction
$\avgDis{C_\vr}$ [Eq.~\eqref{eqCorrelationMeasurement}], is evaluated with data
from Sec.~\ref{secPercolation}. The effect of the zero-cutoff on the correlation
function from Eq.~\eqref{eqCorrelationFunction} is indeed perfectly predicted by
Eq.~\eqref{eqCorrelationDeviation}.


\section{Different versions of the FFM}
\label{secCorrelationOther}
Many different variants of the FFM can be found in the
literature~\cite{Peng1991, Prakash1992, Pang1995, Makse1995, Makse1996,
Ballesteros1999, Ahrens2011,Simon2012}. We want to give a brief overview of the
differences and discuss the effects of some of the implied approximations.

In early works the spectral density is approximated as
\mbox{$S(\vq)=|\vec{q}|^{-(D-a)}$~\cite{Prakash1992}}. The resulting
non-trivial amplitude in the correlation function $C(\vr)=f(D-a)|\vr|^{-a}$ was
shown to vanish for $a\rightarrow D$, in accordance with the picture that the
uncorrelated case should be recovered for short-range correlations ($a>D$).
Still, the desired correlation function could only be produced in a small
region of the system with this approach.

Reference~\cite{Pang1995} follows a similar idea, but directly uses
$C(\vr)=|\vr|^{-a}$. This function diverges at $|\vr|=0$, and hence the authors
interpolate by integrating the function in the corresponding discrete bin
around zero. This works reliably in one dimension but becomes cumbersome in
more dimensions. Moreover, this assigns a non-trivial value to $C(0)$ thus
modifying the variance of the desired Gaussian random variables. 

The most influential works are by Makse et al.~\cite{Makse1995, Makse1996}.
They introduced the correlation function from Eq.~\eqref{eqCorrelationFunction},
allowing them to approach the problem both numerically~\cite{Makse1995} and
(partially) analytically~\cite{Makse1996}.  Their numerical approach is quite
similar to ours, but the analytical one has received far more attention. We have
in fact tried it ourselves~\cite{Fricke2017}, but found that it has many
pitfalls, which we want to briefly discuss here. The idea is to discretize the
Fourier transform of $C(\vr)=(1+\vr^2)^{-a/2}$ for the
infinite continuum, which can be calculated analytically:
\begin{align}
  S(\vq) &=\int_{-\infty}^\infty C(\vr) e^{i\vq\vr} d\vr\nonumber\\
         &=\frac{2\pi^{D/2}}{\Gamma(a/2)}\left(\frac{|\vq|}{2}\right)^{\beta}K_{\beta}(|\vq|),
  \label{eqCorrelationSq}
\end{align}
where $\Gamma$ is Euler's gamma function and $K_\beta(|\vq|)$ is the modified
Bessel function of order $\beta=(a-D)/2$~\footnote{note that there is a typo in
the argument of the Euler gamma function in Ref.~\cite{Makse1996}}. The
variance is recovered by integrating over full continuous space
$\sigma^2_{\mys{}}=C(0)=\frac{1}{2\pi}\int_{-\infty}^{\infty} S(\vq)d\vq=1$.

The next step is to identify $q=\frac{2\pi}{L}\vk$ and map the continuous result to a
discrete lattice by evaluating the function $S(\frac{2\pi}{L}\vk)$ at each lattice site
$\vk$. The first problem here is that $S(\vec{0})$ diverges. This can be
circumvented by evaluating the zero-signal at a shifted frequency, i.e.,
$S_\vec{0}=S(\frac{2\pi}{L}|\vec{k}|_0)$ with $|\vec{k}|_0\in(0,1)$ chosen
``appropriately''~\cite{Makse1996}. With increasing system size the choice
becomes less relevant, and the differences can be expected to vanish in the
infinite-system limit. For finite systems, however, the effect of the parameter
$|\vec{k}|_0$ depends on {the dimension and the strength of the correlations}. In addition,
$|\vec{k}|_0$ has to be adjusted iteratively, rendering the application of the
method rather tedious.

There is another, more severe problem with the discretization, which is relevant
for the mapping to discrete site variables~\cite{Fricke2017} (see
Sec.~\ref{secMapped}).  As we are interested in the asymptotic long-range
scaling behavior, we typically use a fixed lattice spacing of unit length
$\Delta x_i=1$ and consider the limit to infinite system size rather than to the
continuum. Thus, the frequency space is confined to $q_i\in[-\pi,\pi)$, while
the resolution increases with increasing system size. As a consequence, the
variance $\sigma^2_{\mys{}}$ will deviate from one, complicating the mapping
procedure. In fact, we can estimate the deviations via
$\sigma^2_{\mys{}}=\frac{1}{2\pi}\int_{-\pi}^{\pi}
S(\vq)d\vq\leq1=C(0)$. We
numerically verified this but also found additional finite-size scaling
corrections of the form
$\sigma^2_{\mys{},L}=\sigma^2_{\mys{}}+\mathcal{O}(L^{-1})$. These corrections
are inconvenient for finite-size scaling, e.g., for finding the percolation
threshold, because one needs to evaluate the variance $\sigma^2_{\mys{},L}$ in
addition to the parameter $|\vec{k}|_0$ for each {value of the correlation parameter $a$ and each system size}.
By contrast, the method sketched in Sec.~\ref{secCorrelation} is parameter free
and always yields the correct variance $\sigma^2_{\mys{}}$ up to negligible
effects from the zero-cutoff.
\end{appendix}

\bibliography{percolation}

\begin{thebibliography}{49}%
\makeatletter
\providecommand \@ifxundefined [1]{%
 \@ifx{#1\undefined}
}%
\providecommand \@ifnum [1]{%
 \ifnum #1\expandafter \@firstoftwo
 \else \expandafter \@secondoftwo
 \fi
}%
\providecommand \@ifx [1]{%
 \ifx #1\expandafter \@firstoftwo
 \else \expandafter \@secondoftwo
 \fi
}%
\providecommand \natexlab [1]{#1}%
\providecommand \enquote  [1]{``#1''}%
\providecommand \bibnamefont  [1]{#1}%
\providecommand \bibfnamefont [1]{#1}%
\providecommand \citenamefont [1]{#1}%
\providecommand \href@noop [0]{\@secondoftwo}%
\providecommand \href [0]{\begingroup \@sanitize@url \@href}%
\providecommand \@href[1]{\@@startlink{#1}\@@href}%
\providecommand \@@href[1]{\endgroup#1\@@endlink}%
\providecommand \@sanitize@url [0]{\catcode `\\12\catcode `\$12\catcode
  `\&12\catcode `\#12\catcode `\^12\catcode `\_12\catcode `\%12\relax}%
\providecommand \@@startlink[1]{}%
\providecommand \@@endlink[0]{}%
\providecommand \url  [0]{\begingroup\@sanitize@url \@url }%
\providecommand \@url [1]{\endgroup\@href {#1}{\urlprefix }}%
\providecommand \urlprefix  [0]{URL }%
\providecommand \Eprint [0]{\href }%
\providecommand \doibase [0]{http://dx.doi.org/}%
\providecommand \selectlanguage [0]{\@gobble}%
\providecommand \bibinfo  [0]{\@secondoftwo}%
\providecommand \bibfield  [0]{\@secondoftwo}%
\providecommand \translation [1]{[#1]}%
\providecommand \BibitemOpen [0]{}%
\providecommand \bibitemStop [0]{}%
\providecommand \bibitemNoStop [0]{.\EOS\space}%
\providecommand \EOS [0]{\spacefactor3000\relax}%
\providecommand \BibitemShut  [1]{\csname bibitem#1\endcsname}%
\let\auto@bib@innerbib\@empty
\bibitem [{\citenamefont {Pfeifer}\ and\ \citenamefont
  {Avnir}(1983)}]{Avnir1983}%
  \BibitemOpen
  \bibfield  {author} {\bibinfo {author} {\bibfnamefont {P.}~\bibnamefont
  {Pfeifer}}\ and\ \bibinfo {author} {\bibfnamefont {D.}~\bibnamefont
  {Avnir}},\ }\bibfield  {title} {\enquote {\bibinfo {title} {Chemistry in
  noninteger dimensions between two and three. {I}. {F}ractal theory of
  heterogeneous surfaces},}\ }\href@noop {} {\bibfield  {journal} {\bibinfo
  {journal} {J. Chem. Phys.}\ }\textbf {\bibinfo {volume} {79}},\ \bibinfo
  {pages} {3558} (\bibinfo {year} {1983})}\BibitemShut {NoStop}%
\bibitem [{\citenamefont {Avnir}\ \emph {et~al.}(1984)\citenamefont {Avnir},
  \citenamefont {Farin},\ and\ \citenamefont {Pfeifer}}]{Avnir1984}%
  \BibitemOpen
  \bibfield  {author} {\bibinfo {author} {\bibfnamefont {D.}~\bibnamefont
  {Avnir}}, \bibinfo {author} {\bibfnamefont {D.}~\bibnamefont {Farin}}, \ and\
  \bibinfo {author} {\bibfnamefont {P.}~\bibnamefont {Pfeifer}},\ }\bibfield
  {title} {\enquote {\bibinfo {title} {Molecular fractal surfaces},}\
  }\href@noop {} {\bibfield  {journal} {\bibinfo  {journal} {Nature}\ }\textbf
  {\bibinfo {volume} {308}},\ \bibinfo {pages} {261} (\bibinfo {year}
  {1984})}\BibitemShut {NoStop}%
\bibitem [{\citenamefont {Dorogovtsev}(1980)}]{Dorogovtsev1980}%
  \BibitemOpen
  \bibfield  {author} {\bibinfo {author} {\bibfnamefont {S.~N.}\ \bibnamefont
  {Dorogovtsev}},\ }\bibfield  {title} {\enquote {\bibinfo {title} {Critical
  exponents of magnets with lengthy defects},}\ }\href {\doibase
  http://dx.doi.org/10.1016/0375-9601(80)90604-0} {\bibfield  {journal}
  {\bibinfo  {journal} {Phys. Lett. A}\ }\textbf {\bibinfo {volume} {76}},\
  \bibinfo {pages} {169} (\bibinfo {year} {1980})}\BibitemShut {NoStop}%
\bibitem [{\citenamefont {Yamazaki}\ \emph {et~al.}(1986)\citenamefont
  {Yamazaki}, \citenamefont {Holz}, \citenamefont {Ochiai},\ and\ \citenamefont
  {Fukuda}}]{Yamazaki1986a}%
  \BibitemOpen
  \bibfield  {author} {\bibinfo {author} {\bibfnamefont {Y.}~\bibnamefont
  {Yamazaki}}, \bibinfo {author} {\bibfnamefont {A.}~\bibnamefont {Holz}},
  \bibinfo {author} {\bibfnamefont {M.}~\bibnamefont {Ochiai}}, \ and\ \bibinfo
  {author} {\bibfnamefont {Y.}~\bibnamefont {Fukuda}},\ }\bibfield  {title}
  {\enquote {\bibinfo {title} {Static and dynamic critical behavior of
  extended-defect {$N$}-component systems in cubic anisotropic crystals},}\
  }\href {\doibase 10.1103/PhysRevB.33.3460} {\bibfield  {journal} {\bibinfo
  {journal} {Phys. Rev. B}\ }\textbf {\bibinfo {volume} {33}},\ \bibinfo
  {pages} {3460} (\bibinfo {year} {1986})}\BibitemShut {NoStop}%
\bibitem [{\citenamefont {Bouchaud}\ and\ \citenamefont
  {Georges}(1990)}]{Bouchaud1990}%
  \BibitemOpen
  \bibfield  {author} {\bibinfo {author} {\bibfnamefont {J.-P.}\ \bibnamefont
  {Bouchaud}}\ and\ \bibinfo {author} {\bibfnamefont {A.}~\bibnamefont
  {Georges}},\ }\bibfield  {title} {\enquote {\bibinfo {title} {Anomalous
  diffusion in disordered media: {S}tatistical mechanisms, models and physical
  applications},}\ }\href@noop {} {\bibfield  {journal} {\bibinfo  {journal}
  {Phys. Rep.}\ }\textbf {\bibinfo {volume} {195}},\ \bibinfo {pages} {127}
  (\bibinfo {year} {1990})}\BibitemShut {NoStop}%
\bibitem [{\citenamefont {Malek}\ and\ \citenamefont
  {Coppens}(2001)}]{Malek2001}%
  \BibitemOpen
  \bibfield  {author} {\bibinfo {author} {\bibfnamefont {K.}~\bibnamefont
  {Malek}}\ and\ \bibinfo {author} {\bibfnamefont {M.-O.}\ \bibnamefont
  {Coppens}},\ }\bibfield  {title} {\enquote {\bibinfo {title} {Effects of
  surface roughness on self- and transport diffusion in porous media in the
  {K}nudsen regime},}\ }\href {\doibase 10.1103/PhysRevLett.87.125505}
  {\bibfield  {journal} {\bibinfo  {journal} {Phys. Rev. Lett.}\ }\textbf
  {\bibinfo {volume} {87}},\ \bibinfo {pages} {125505} (\bibinfo {year}
  {2001})}\BibitemShut {NoStop}%
\bibitem [{\citenamefont {Foulaadvand}\ and\ \citenamefont
  {Sadrara}(2015)}]{Foulaadvand2015}%
  \BibitemOpen
  \bibfield  {author} {\bibinfo {author} {\bibfnamefont {M.~E.}\ \bibnamefont
  {Foulaadvand}}\ and\ \bibinfo {author} {\bibfnamefont {M.}~\bibnamefont
  {Sadrara}},\ }\bibfield  {title} {\enquote {\bibinfo {title} {Dynamics of a
  rigid rod in a disordered medium with long-range spatial correlation},}\
  }\href@noop {} {\bibfield  {journal} {\bibinfo  {journal} {Phys. Rev. E}\
  }\textbf {\bibinfo {volume} {91}},\ \bibinfo {pages} {012122} (\bibinfo
  {year} {2015})}\BibitemShut {NoStop}%
\bibitem [{\citenamefont {Goychuk}\ \emph {et~al.}(2017)\citenamefont
  {Goychuk}, \citenamefont {Kharchenko},\ and\ \citenamefont
  {Metzler}}]{Goychuk2017}%
  \BibitemOpen
  \bibfield  {author} {\bibinfo {author} {\bibfnamefont {I.}~\bibnamefont
  {Goychuk}}, \bibinfo {author} {\bibfnamefont {V.~O.}\ \bibnamefont
  {Kharchenko}}, \ and\ \bibinfo {author} {\bibfnamefont {R.}~\bibnamefont
  {Metzler}},\ }\bibfield  {title} {\enquote {\bibinfo {title} {Persistent
  {S}inai type diffusion in {G}aussian random potentials with decaying spatial
  correlations},}\ }\href@noop {} {\bibfield  {journal} {\bibinfo  {journal}
  {Phys. Rev. E}\ }\textbf {\bibinfo {volume} {96}},\ \bibinfo {pages} {052134}
  (\bibinfo {year} {2017})}\BibitemShut {NoStop}%
\bibitem [{\citenamefont {Dullien}(1979)}]{Dullien1979}%
  \BibitemOpen
  \bibfield  {author} {\bibinfo {author} {\bibfnamefont {F.~A.~L.}\
  \bibnamefont {Dullien}},\ }\href@noop {} {\emph {\bibinfo {title} {Porous
  {Media}: {Fluid} {Transport} and {Pore} {Structure}}}}\ (\bibinfo
  {publisher} {Academic Press},\ \bibinfo {address} {New York},\ \bibinfo
  {year} {1979})\BibitemShut {NoStop}%
\bibitem [{\citenamefont {Sahimi}(1995)}]{Sahimi1995}%
  \BibitemOpen
  \bibfield  {author} {\bibinfo {author} {\bibfnamefont {M.}~\bibnamefont
  {Sahimi}},\ }\href@noop {} {\emph {\bibinfo {title} {Flow and {T}ransport in
  {P}orous {M}edia and {F}ractured {R}ock}}}\ (\bibinfo  {publisher} {VCH},\
  \bibinfo {address} {Weinheim},\ \bibinfo {year} {1995})\BibitemShut {NoStop}%
\bibitem [{\citenamefont {Skinner}\ \emph {et~al.}(2013)\citenamefont
  {Skinner}, \citenamefont {Schnyder}, \citenamefont {Aarts}, \citenamefont
  {Horbach},\ and\ \citenamefont {Dullens}}]{Skinner2013}%
  \BibitemOpen
  \bibfield  {author} {\bibinfo {author} {\bibfnamefont {T.~O.~E.}\
  \bibnamefont {Skinner}}, \bibinfo {author} {\bibfnamefont {S.~K.}\
  \bibnamefont {Schnyder}}, \bibinfo {author} {\bibfnamefont {D.~G.~A.~L.}\
  \bibnamefont {Aarts}}, \bibinfo {author} {\bibfnamefont {J.}~\bibnamefont
  {Horbach}}, \ and\ \bibinfo {author} {\bibfnamefont {R.~P.~A.}\ \bibnamefont
  {Dullens}},\ }\bibfield  {title} {\enquote {\bibinfo {title} {Localization
  dynamics of fluids in random confinement},}\ }\href@noop {} {\bibfield
  {journal} {\bibinfo  {journal} {Phys. Rev. Lett.}\ }\textbf {\bibinfo
  {volume} {111}},\ \bibinfo {pages} {128301} (\bibinfo {year}
  {2013})}\BibitemShut {NoStop}%
\bibitem [{\citenamefont {Spanner}\ \emph {et~al.}(2016)\citenamefont
  {Spanner}, \citenamefont {H{\"o}fling}, \citenamefont {Kapfer}, \citenamefont
  {Mecke}, \citenamefont {Schr{\"o}der-Turk},\ and\ \citenamefont
  {Franosch}}]{Spanner2016}%
  \BibitemOpen
  \bibfield  {author} {\bibinfo {author} {\bibfnamefont {M.}~\bibnamefont
  {Spanner}}, \bibinfo {author} {\bibfnamefont {F.}~\bibnamefont
  {H{\"o}fling}}, \bibinfo {author} {\bibfnamefont {S.~C.}\ \bibnamefont
  {Kapfer}}, \bibinfo {author} {\bibfnamefont {K.~R.}\ \bibnamefont {Mecke}},
  \bibinfo {author} {\bibfnamefont {G.~E.}\ \bibnamefont {Schr{\"o}der-Turk}},
  \ and\ \bibinfo {author} {\bibfnamefont {T.}~\bibnamefont {Franosch}},\
  }\bibfield  {title} {\enquote {\bibinfo {title} {Splitting of the
  universality class of anomalous transport in crowded media},}\ }\href@noop {}
  {\bibfield  {journal} {\bibinfo  {journal} {Phys. Rev. Lett.}\ }\textbf
  {\bibinfo {volume} {116}},\ \bibinfo {pages} {060601} (\bibinfo {year}
  {2016})}\BibitemShut {NoStop}%
\bibitem [{\citenamefont {Bancaud}\ \emph {et~al.}(2012)\citenamefont
  {Bancaud}, \citenamefont {Lavelle}, \citenamefont {Huet},\ and\ \citenamefont
  {Ellenberg}}]{Bancaud2012}%
  \BibitemOpen
  \bibfield  {author} {\bibinfo {author} {\bibfnamefont {A.}~\bibnamefont
  {Bancaud}}, \bibinfo {author} {\bibfnamefont {C.}~\bibnamefont {Lavelle}},
  \bibinfo {author} {\bibfnamefont {S.}~\bibnamefont {Huet}}, \ and\ \bibinfo
  {author} {\bibfnamefont {J.}~\bibnamefont {Ellenberg}},\ }\bibfield  {title}
  {\enquote {\bibinfo {title} {A fractal model for nuclear organization:
  Current evidence and biological implications},}\ }\href {\doibase
  10.1093/nar/gks586} {\bibfield  {journal} {\bibinfo  {journal} {Nucleic Acids
  Res.}\ }\textbf {\bibinfo {volume} {40}},\ \bibinfo {pages} {8783} (\bibinfo
  {year} {2012})}\BibitemShut {NoStop}%
\bibitem [{\citenamefont {H\"ofling}\ and\ \citenamefont
  {Franosch}(2013)}]{Hoefling-Franosch2013}%
  \BibitemOpen
  \bibfield  {author} {\bibinfo {author} {\bibfnamefont {F.}~\bibnamefont
  {H\"ofling}}\ and\ \bibinfo {author} {\bibfnamefont {T.}~\bibnamefont
  {Franosch}},\ }\bibfield  {title} {\enquote {\bibinfo {title} {Anomalous
  transport in the crowded world of biological cells},}\ }\href@noop {}
  {\bibfield  {journal} {\bibinfo  {journal} {Rep. Prog. Phys.}\ }\textbf
  {\bibinfo {volume} {76}},\ \bibinfo {pages} {046602} (\bibinfo {year}
  {2013})}\BibitemShut {NoStop}%
\bibitem [{\citenamefont {Stauffer}\ and\ \citenamefont
  {Aharony}(1992)}]{Stauffer1992}%
  \BibitemOpen
  \bibfield  {author} {\bibinfo {author} {\bibfnamefont {D.}~\bibnamefont
  {Stauffer}}\ and\ \bibinfo {author} {\bibfnamefont {A.}~\bibnamefont
  {Aharony}},\ }\href@noop {} {\emph {\bibinfo {title} {Introduction to
  Percolation Theory}}}\ (\bibinfo  {publisher} {Taylor and Francis},\ \bibinfo
  {address} {London},\ \bibinfo {year} {1992})\BibitemShut {NoStop}%
\bibitem [{\citenamefont {Weinrib}\ and\ \citenamefont
  {Halperin}(1983)}]{Weinrib1983}%
  \BibitemOpen
  \bibfield  {author} {\bibinfo {author} {\bibfnamefont {A.}~\bibnamefont
  {Weinrib}}\ and\ \bibinfo {author} {\bibfnamefont {B.~I.}\ \bibnamefont
  {Halperin}},\ }\bibfield  {title} {\enquote {\bibinfo {title} {Critical
  phenomena in systems with long-range-correlated quenched disorder},}\
  }\href@noop {} {\bibfield  {journal} {\bibinfo  {journal} {Phys. Rev. B}\
  }\textbf {\bibinfo {volume} {27}},\ \bibinfo {pages} {413} (\bibinfo {year}
  {1983})}\BibitemShut {NoStop}%
\bibitem [{\citenamefont {Weinrib}(1984)}]{Weinrib1984}%
  \BibitemOpen
  \bibfield  {author} {\bibinfo {author} {\bibfnamefont {A.}~\bibnamefont
  {Weinrib}},\ }\bibfield  {title} {\enquote {\bibinfo {title} {Long-range
  correlated percolation},}\ }\href {\doibase 10.1103/PhysRevB.29.387}
  {\bibfield  {journal} {\bibinfo  {journal} {Phys. Rev. B}\ }\textbf {\bibinfo
  {volume} {29}},\ \bibinfo {pages} {387} (\bibinfo {year} {1984})}\BibitemShut
  {NoStop}%
\bibitem [{\citenamefont {Harris}(1974)}]{Harris1974}%
  \BibitemOpen
  \bibfield  {author} {\bibinfo {author} {\bibfnamefont {A.~B.}\ \bibnamefont
  {Harris}},\ }\bibfield  {title} {\enquote {\bibinfo {title} {Effect of random
  defects on the critical behaviour of {I}sing models},}\ }\href@noop {}
  {\bibfield  {journal} {\bibinfo  {journal} {J. Phys. C}\ }\textbf {\bibinfo
  {volume} {7}},\ \bibinfo {pages} {1671} (\bibinfo {year} {1974})}\BibitemShut
  {NoStop}%
\bibitem [{\citenamefont {Prudnikov}\ and\ \citenamefont
  {Fedorenko}(1999)}]{Prudnikov1999}%
  \BibitemOpen
  \bibfield  {author} {\bibinfo {author} {\bibfnamefont {V.~V.}\ \bibnamefont
  {Prudnikov}}\ and\ \bibinfo {author} {\bibfnamefont {A.~A.}\ \bibnamefont
  {Fedorenko}},\ }\bibfield  {title} {\enquote {\bibinfo {title} {Critical
  behaviour of {3D} systems with long-range-correlated quenched disorder},}\
  }\href@noop {} {\bibfield  {journal} {\bibinfo  {journal} {J. Phys. A: Math.
  Gen.}\ }\textbf {\bibinfo {volume} {32}},\ \bibinfo {pages} {L399} (\bibinfo
  {year} {1999})}\BibitemShut {NoStop}%
\bibitem [{\citenamefont {Prudnikov}\ \emph {et~al.}(2000)\citenamefont
  {Prudnikov}, \citenamefont {Prudnikov},\ and\ \citenamefont
  {Fedorenko}}]{Prudnikov2000}%
  \BibitemOpen
  \bibfield  {author} {\bibinfo {author} {\bibfnamefont {V.~V.}\ \bibnamefont
  {Prudnikov}}, \bibinfo {author} {\bibfnamefont {P.~V.}\ \bibnamefont
  {Prudnikov}}, \ and\ \bibinfo {author} {\bibfnamefont {A.~A.}\ \bibnamefont
  {Fedorenko}},\ }\bibfield  {title} {\enquote {\bibinfo {title} {Field-theory
  approach to critical behavior of systems with long-range-correlated
  defects},}\ }\href@noop {} {\bibfield  {journal} {\bibinfo  {journal} {Phys.
  Rev. B}\ }\textbf {\bibinfo {volume} {62}},\ \bibinfo {pages} {8777}
  (\bibinfo {year} {2000})}\BibitemShut {NoStop}%
\bibitem [{\citenamefont {Schrenk}\ \emph {et~al.}(2013)\citenamefont
  {Schrenk}, \citenamefont {Pos\'e}, \citenamefont {Kranz}, \citenamefont {van
  Kessenich}, \citenamefont {Ara\'ujo},\ and\ \citenamefont
  {Herrmann}}]{Schrenk2013}%
  \BibitemOpen
  \bibfield  {author} {\bibinfo {author} {\bibfnamefont {K.~J.}\ \bibnamefont
  {Schrenk}}, \bibinfo {author} {\bibfnamefont {N.}~\bibnamefont {Pos\'e}},
  \bibinfo {author} {\bibfnamefont {J.~J.}\ \bibnamefont {Kranz}}, \bibinfo
  {author} {\bibfnamefont {L.~V.~M.}\ \bibnamefont {van Kessenich}}, \bibinfo
  {author} {\bibfnamefont {N.~A.~M.}\ \bibnamefont {Ara\'ujo}}, \ and\ \bibinfo
  {author} {\bibfnamefont {H.~J.}\ \bibnamefont {Herrmann}},\ }\bibfield
  {title} {\enquote {\bibinfo {title} {Percolation with long-range correlated
  disorder},}\ }\href {\doibase 10.1103/PhysRevE.88.052102} {\bibfield
  {journal} {\bibinfo  {journal} {Phys. Rev. E}\ }\textbf {\bibinfo {volume}
  {88}},\ \bibinfo {pages} {052102} (\bibinfo {year} {2013})}\BibitemShut
  {NoStop}%
\bibitem [{\citenamefont {Prakash}\ \emph {et~al.}(1992)\citenamefont
  {Prakash}, \citenamefont {Havlin}, \citenamefont {Schwartz},\ and\
  \citenamefont {Stanley}}]{Prakash1992}%
  \BibitemOpen
  \bibfield  {author} {\bibinfo {author} {\bibfnamefont {S.}~\bibnamefont
  {Prakash}}, \bibinfo {author} {\bibfnamefont {S.}~\bibnamefont {Havlin}},
  \bibinfo {author} {\bibfnamefont {M.}~\bibnamefont {Schwartz}}, \ and\
  \bibinfo {author} {\bibfnamefont {H.~E.}\ \bibnamefont {Stanley}},\
  }\bibfield  {title} {\enquote {\bibinfo {title} {Structural and dynamical
  properties of long-range correlated percolation},}\ }\href {\doibase
  10.1103/PhysRevA.46.R1724} {\bibfield  {journal} {\bibinfo  {journal} {Phys.
  Rev. A}\ }\textbf {\bibinfo {volume} {46}},\ \bibinfo {pages} {R1724}
  (\bibinfo {year} {1992})}\BibitemShut {NoStop}%
\bibitem [{\citenamefont {Makse}\ \emph {et~al.}(1998)\citenamefont {Makse},
  \citenamefont {Andrade~Jr.}, \citenamefont {Batty}, \citenamefont {Havlin},\
  and\ \citenamefont {Stanley}}]{Makse1998}%
  \BibitemOpen
  \bibfield  {author} {\bibinfo {author} {\bibfnamefont {H.~A.}\ \bibnamefont
  {Makse}}, \bibinfo {author} {\bibfnamefont {J.~S.}\ \bibnamefont
  {Andrade~Jr.}}, \bibinfo {author} {\bibfnamefont {M.}~\bibnamefont {Batty}},
  \bibinfo {author} {\bibfnamefont {S.}~\bibnamefont {Havlin}}, \ and\ \bibinfo
  {author} {\bibfnamefont {H.~E.}\ \bibnamefont {Stanley}},\ }\bibfield
  {title} {\enquote {\bibinfo {title} {Modeling urban growth patterns with
  correlated percolation},}\ }\href@noop {} {\bibfield  {journal} {\bibinfo
  {journal} {Phys. Rev. E}\ }\textbf {\bibinfo {volume} {58}},\ \bibinfo
  {pages} {7054} (\bibinfo {year} {1998})}\BibitemShut {NoStop}%
\bibitem [{\citenamefont {Saupe}(1988)}]{Saupe1988}%
  \BibitemOpen
  \bibfield  {author} {\bibinfo {author} {\bibfnamefont {D.}~\bibnamefont
  {Saupe}},\ }\bibfield  {title} {\enquote {\bibinfo {title} {Algorithms for
  random fractals},}\ }in\ \href@noop {} {\emph {\bibinfo {booktitle} {The
  Science of Fractal Images}}},\ \bibinfo {editor} {edited by\ \bibinfo
  {editor} {\bibfnamefont {H.-O.}\ \bibnamefont {Peitgen}}\ and\ \bibinfo
  {editor} {\bibfnamefont {D.}~\bibnamefont {Saupe}}}\ (\bibinfo  {publisher}
  {Springer},\ \bibinfo {address} {New York},\ \bibinfo {year} {1988})\ pp.\
  \bibinfo {pages} {71--136}\BibitemShut {NoStop}%
\bibitem [{\citenamefont {Peng}\ \emph {et~al.}(1991)\citenamefont {Peng},
  \citenamefont {Havlin}, \citenamefont {Schwartz},\ and\ \citenamefont
  {Stanley}}]{Peng1991}%
  \BibitemOpen
  \bibfield  {author} {\bibinfo {author} {\bibfnamefont {C.-K.}\ \bibnamefont
  {Peng}}, \bibinfo {author} {\bibfnamefont {S.}~\bibnamefont {Havlin}},
  \bibinfo {author} {\bibfnamefont {M.}~\bibnamefont {Schwartz}}, \ and\
  \bibinfo {author} {\bibfnamefont {H.~E.}\ \bibnamefont {Stanley}},\
  }\bibfield  {title} {\enquote {\bibinfo {title} {Directed-polymer and
  ballistic-deposition growth with correlated noise},}\ }\href {\doibase
  10.1103/PhysRevA.44.R2239} {\bibfield  {journal} {\bibinfo  {journal} {Phys.
  Rev. A}\ }\textbf {\bibinfo {volume} {44}},\ \bibinfo {pages} {R2239}
  (\bibinfo {year} {1991})}\BibitemShut {NoStop}%
\bibitem [{\citenamefont {Makse}\ \emph
  {et~al.}(1995{\natexlab{a}})\citenamefont {Makse}, \citenamefont {Havlin},
  \citenamefont {Stanley},\ and\ \citenamefont {Schwartz}}]{Makse1995}%
  \BibitemOpen
  \bibfield  {author} {\bibinfo {author} {\bibfnamefont {H.~A.}\ \bibnamefont
  {Makse}}, \bibinfo {author} {\bibfnamefont {S.}~\bibnamefont {Havlin}},
  \bibinfo {author} {\bibfnamefont {H.~E.}\ \bibnamefont {Stanley}}, \ and\
  \bibinfo {author} {\bibfnamefont {M.}~\bibnamefont {Schwartz}},\ }\bibfield
  {title} {\enquote {\bibinfo {title} {Novel method for generating long-range
  correlations},}\ }\href@noop {} {\bibfield  {journal} {\bibinfo  {journal}
  {Chaos, Solitons \& Fractals}\ }\textbf {\bibinfo {volume} {6}},\ \bibinfo
  {pages} {295} (\bibinfo {year} {1995}{\natexlab{a}})}\BibitemShut {NoStop}%
\bibitem [{\citenamefont {Pang}\ \emph {et~al.}(1995)\citenamefont {Pang},
  \citenamefont {Yu},\ and\ \citenamefont {Halpin-Healy}}]{Pang1995}%
  \BibitemOpen
  \bibfield  {author} {\bibinfo {author} {\bibfnamefont {N.-N.}\ \bibnamefont
  {Pang}}, \bibinfo {author} {\bibfnamefont {Y.-K.}\ \bibnamefont {Yu}}, \ and\
  \bibinfo {author} {\bibfnamefont {T.}~\bibnamefont {Halpin-Healy}},\
  }\bibfield  {title} {\enquote {\bibinfo {title} {Interfacial kinetic
  roughening with correlated noise},}\ }\href@noop {} {\bibfield  {journal}
  {\bibinfo  {journal} {Phys. Rev. E}\ }\textbf {\bibinfo {volume} {52}},\
  \bibinfo {pages} {3224} (\bibinfo {year} {1995})}\BibitemShut {NoStop}%
\bibitem [{\citenamefont {Makse}\ \emph {et~al.}(1996)\citenamefont {Makse},
  \citenamefont {Havlin}, \citenamefont {Schwartz},\ and\ \citenamefont
  {Stanley}}]{Makse1996}%
  \BibitemOpen
  \bibfield  {author} {\bibinfo {author} {\bibfnamefont {H.~A.}\ \bibnamefont
  {Makse}}, \bibinfo {author} {\bibfnamefont {S.}~\bibnamefont {Havlin}},
  \bibinfo {author} {\bibfnamefont {M.}~\bibnamefont {Schwartz}}, \ and\
  \bibinfo {author} {\bibfnamefont {H.~E.}\ \bibnamefont {Stanley}},\
  }\bibfield  {title} {\enquote {\bibinfo {title} {Method for generating
  long-range correlations for large systems},}\ }\href {\doibase
  10.1103/PhysRevE.53.5445} {\bibfield  {journal} {\bibinfo  {journal} {Phys.
  Rev. E}\ }\textbf {\bibinfo {volume} {53}},\ \bibinfo {pages} {5445}
  (\bibinfo {year} {1996})}\BibitemShut {NoStop}%
\bibitem [{\citenamefont {Ballesteros}\ and\ \citenamefont
  {Parisi}(1999)}]{Ballesteros1999}%
  \BibitemOpen
  \bibfield  {author} {\bibinfo {author} {\bibfnamefont {H.~G.}\ \bibnamefont
  {Ballesteros}}\ and\ \bibinfo {author} {\bibfnamefont {G.}~\bibnamefont
  {Parisi}},\ }\bibfield  {title} {\enquote {\bibinfo {title} {Site-diluted
  three-dimensional {I}sing model with long-range correlated disorder},}\
  }\href@noop {} {\bibfield  {journal} {\bibinfo  {journal} {Phys. Rev. B}\
  }\textbf {\bibinfo {volume} {60}},\ \bibinfo {pages} {12912} (\bibinfo {year}
  {1999})}\BibitemShut {NoStop}%
\bibitem [{\citenamefont {Ahrens}\ and\ \citenamefont
  {Hartmann}(2011)}]{Ahrens2011}%
  \BibitemOpen
  \bibfield  {author} {\bibinfo {author} {\bibfnamefont {B.}~\bibnamefont
  {Ahrens}}\ and\ \bibinfo {author} {\bibfnamefont {A.~K.}\ \bibnamefont
  {Hartmann}},\ }\bibfield  {title} {\enquote {\bibinfo {title} {Critical
  behavior of the random-field {I}sing magnet with long-range correlated
  disorder},}\ }\href@noop {} {\bibfield  {journal} {\bibinfo  {journal} {Phys.
  Rev. B}\ }\textbf {\bibinfo {volume} {84}},\ \bibinfo {pages} {144202}
  (\bibinfo {year} {2011})}\BibitemShut {NoStop}%
\bibitem [{\citenamefont {Simon}\ \emph {et~al.}(2012)\citenamefont {Simon},
  \citenamefont {Sancho},\ and\ \citenamefont {Lacasta}}]{Simon2012}%
  \BibitemOpen
  \bibfield  {author} {\bibinfo {author} {\bibfnamefont {M.~S.}\ \bibnamefont
  {Simon}}, \bibinfo {author} {\bibfnamefont {J.~M.}\ \bibnamefont {Sancho}}, \
  and\ \bibinfo {author} {\bibfnamefont {A.~M.}\ \bibnamefont {Lacasta}},\
  }\bibfield  {title} {\enquote {\bibinfo {title} {On generating random
  potentials},}\ }\href@noop {} {\bibfield  {journal} {\bibinfo  {journal}
  {Fluct. Noise Lett.}\ }\textbf {\bibinfo {volume} {11}},\ \bibinfo {pages}
  {1250026} (\bibinfo {year} {2012})}\BibitemShut {NoStop}%
\bibitem [{Note1()}]{Note1}%
  \BibitemOpen
  \bibinfo {note} {Our code (C++) is available at \protect \url
  {github.com/CQT-Leipzig/correlated_disorder}}\BibitemShut {NoStop}%
\bibitem [{\citenamefont {Press}\ \emph {et~al.}(2007)\citenamefont {Press},
  \citenamefont {Teukolsky}, \citenamefont {Vetterling},\ and\ \citenamefont
  {Flannery}}]{NumericalRecipiesFFT}%
  \BibitemOpen
  \bibfield  {author} {\bibinfo {author} {\bibfnamefont {W.~H.}\ \bibnamefont
  {Press}}, \bibinfo {author} {\bibfnamefont {S.~A.}\ \bibnamefont
  {Teukolsky}}, \bibinfo {author} {\bibfnamefont {W.~T.}\ \bibnamefont
  {Vetterling}}, \ and\ \bibinfo {author} {\bibfnamefont {B.~P.}\ \bibnamefont
  {Flannery}},\ }\href@noop {} {\emph {\bibinfo {title} {Numerical {R}ecipes
  3rd edition: The {A}rt of {S}cientific {C}omputing}}}\ (\bibinfo  {publisher}
  {Cambridge University Press},\ \bibinfo {address} {Cambridge},\ \bibinfo
  {year} {2007})\BibitemShut {NoStop}%
\bibitem [{\citenamefont {Wiseman}\ and\ \citenamefont
  {Domany}(1998)}]{Wiseman1998}%
  \BibitemOpen
  \bibfield  {author} {\bibinfo {author} {\bibfnamefont {S.}~\bibnamefont
  {Wiseman}}\ and\ \bibinfo {author} {\bibfnamefont {E.}~\bibnamefont
  {Domany}},\ }\bibfield  {title} {\enquote {\bibinfo {title} {Self-averaging,
  distribution of pseudocritical temperatures, and finite size scaling in
  critical disordered systems},}\ }\href@noop {} {\bibfield  {journal}
  {\bibinfo  {journal} {Phys. Rev. E}\ }\textbf {\bibinfo {volume} {58}},\
  \bibinfo {pages} {2938} (\bibinfo {year} {1998})}\BibitemShut {NoStop}%
\bibitem [{\citenamefont {Newman}\ and\ \citenamefont
  {Ziff}(2001)}]{Newman2001}%
  \BibitemOpen
  \bibfield  {author} {\bibinfo {author} {\bibfnamefont {M.~E.~J.}\
  \bibnamefont {Newman}}\ and\ \bibinfo {author} {\bibfnamefont {R.~M.}\
  \bibnamefont {Ziff}},\ }\bibfield  {title} {\enquote {\bibinfo {title} {Fast
  {M}onte {C}arlo algorithm for site or bond percolation},}\ }\href {\doibase
  10.1103/PhysRevE.64.016706} {\bibfield  {journal} {\bibinfo  {journal} {Phys.
  Rev. E}\ }\textbf {\bibinfo {volume} {64}},\ \bibinfo {pages} {016706}
  (\bibinfo {year} {2001})}\BibitemShut {NoStop}%
\bibitem [{\citenamefont {Nienhuis}(1984)}]{Nienhuis1984}%
  \BibitemOpen
  \bibfield  {author} {\bibinfo {author} {\bibfnamefont {B.}~\bibnamefont
  {Nienhuis}},\ }\bibfield  {title} {\enquote {\bibinfo {title} {Critical
  behavior of two-dimensional spin models and charge asymmetry in the {C}oulomb
  gas},}\ }\href {\doibase 10.1007/BF01009437} {\bibfield  {journal} {\bibinfo
  {journal} {J. Stat. Phys.}\ }\textbf {\bibinfo {volume} {34}},\ \bibinfo
  {pages} {731} (\bibinfo {year} {1984})}\BibitemShut {NoStop}%
\bibitem [{\citenamefont {Ziff}(1992)}]{Ziff1994}%
  \BibitemOpen
  \bibfield  {author} {\bibinfo {author} {\bibfnamefont {R.~M.}\ \bibnamefont
  {Ziff}},\ }\bibfield  {title} {\enquote {\bibinfo {title} {Spanning
  probability in {2D} percolation},}\ }\href@noop {} {\bibfield  {journal}
  {\bibinfo  {journal} {Phys. Rev. Lett.}\ }\textbf {\bibinfo {volume} {69}},\
  \bibinfo {pages} {2670} (\bibinfo {year} {1992})}\BibitemShut {NoStop}%
\bibitem [{\citenamefont {Fricke}\ \emph {et~al.}(2017)\citenamefont {Fricke},
  \citenamefont {Zierenberg}, \citenamefont {Marenz}, \citenamefont {Spitzner},
  \citenamefont {Blavatska},\ and\ \citenamefont {Janke}}]{Fricke2017}%
  \BibitemOpen
  \bibfield  {author} {\bibinfo {author} {\bibfnamefont {N.}~\bibnamefont
  {Fricke}}, \bibinfo {author} {\bibfnamefont {J.}~\bibnamefont {Zierenberg}},
  \bibinfo {author} {\bibfnamefont {M.}~\bibnamefont {Marenz}}, \bibinfo
  {author} {\bibfnamefont {F.~P.}\ \bibnamefont {Spitzner}}, \bibinfo {author}
  {\bibfnamefont {V.}~\bibnamefont {Blavatska}}, \ and\ \bibinfo {author}
  {\bibfnamefont {W.}~\bibnamefont {Janke}},\ }\bibfield  {title} {\enquote
  {\bibinfo {title} {Scaling laws for random walks in long-range correlated
  disordered media},}\ }\href@noop {} {\bibfield  {journal} {\bibinfo
  {journal} {Condens. Matter Phys.}\ }\textbf {\bibinfo {volume} {20}},\
  \bibinfo {pages} {13004} (\bibinfo {year} {2017})}\BibitemShut {NoStop}%
\bibitem [{\citenamefont {Herrmann}\ and\ \citenamefont
  {Stauffer}(1983)}]{Herrmann1983}%
  \BibitemOpen
  \bibfield  {author} {\bibinfo {author} {\bibfnamefont {H.~J.}\ \bibnamefont
  {Herrmann}}\ and\ \bibinfo {author} {\bibfnamefont {D.}~\bibnamefont
  {Stauffer}},\ }\bibfield  {title} {\enquote {\bibinfo {title} {Corrections to
  scaling and finite size effects},}\ }\href@noop {} {\bibfield  {journal}
  {\bibinfo  {journal} {Phys. Lett. A}\ }\textbf {\bibinfo {volume} {100}},\
  \bibinfo {pages} {366} (\bibinfo {year} {1983})}\BibitemShut {NoStop}%
\bibitem [{\citenamefont {Wang}\ \emph {et~al.}(2013)\citenamefont {Wang},
  \citenamefont {Zhou}, \citenamefont {Zhang}, \citenamefont {Garoni},\ and\
  \citenamefont {Deng}}]{Wang2013}%
  \BibitemOpen
  \bibfield  {author} {\bibinfo {author} {\bibfnamefont {J.}~\bibnamefont
  {Wang}}, \bibinfo {author} {\bibfnamefont {Z.}~\bibnamefont {Zhou}}, \bibinfo
  {author} {\bibfnamefont {W.}~\bibnamefont {Zhang}}, \bibinfo {author}
  {\bibfnamefont {T.~M.}\ \bibnamefont {Garoni}}, \ and\ \bibinfo {author}
  {\bibfnamefont {Y.}~\bibnamefont {Deng}},\ }\bibfield  {title} {\enquote
  {\bibinfo {title} {Bond and site percolation in three dimensions},}\
  }\href@noop {} {\bibfield  {journal} {\bibinfo  {journal} {Phys. Rev. E}\
  }\textbf {\bibinfo {volume} {87}},\ \bibinfo {pages} {052107} (\bibinfo
  {year} {2013})}\BibitemShut {NoStop}%
\bibitem [{\citenamefont {Xu}\ \emph {et~al.}(2014)\citenamefont {Xu},
  \citenamefont {Wang}, \citenamefont {Lv},\ and\ \citenamefont
  {Deng}}]{Xu2014}%
  \BibitemOpen
  \bibfield  {author} {\bibinfo {author} {\bibfnamefont {X}~\bibnamefont {Xu}},
  \bibinfo {author} {\bibfnamefont {H}~\bibnamefont {Wang}}, \bibinfo {author}
  {\bibfnamefont {J.-P.}\ \bibnamefont {Lv}}, \ and\ \bibinfo {author}
  {\bibfnamefont {Y.}~\bibnamefont {Deng}},\ }\bibfield  {title} {\enquote
  {\bibinfo {title} {Simultaneous analysis of three-dimensional percolation
  models},}\ }\href@noop {} {\bibfield  {journal} {\bibinfo  {journal} {Front.
  Phys.}\ }\textbf {\bibinfo {volume} {9}},\ \bibinfo {pages} {113} (\bibinfo
  {year} {2014})}\BibitemShut {NoStop}%
\bibitem [{\citenamefont {{Hu}}\ \emph {et~al.}(2014)\citenamefont {{Hu}},
  \citenamefont {{Bl{\"o}te}}, \citenamefont {{Ziff}},\ and\ \citenamefont
  {{Deng}}}]{Hu2014}%
  \BibitemOpen
  \bibfield  {author} {\bibinfo {author} {\bibfnamefont {H.}~\bibnamefont
  {{Hu}}}, \bibinfo {author} {\bibfnamefont {H.~W.~J.}\ \bibnamefont
  {{Bl{\"o}te}}}, \bibinfo {author} {\bibfnamefont {R.~M.}\ \bibnamefont
  {{Ziff}}}, \ and\ \bibinfo {author} {\bibfnamefont {Y.}~\bibnamefont
  {{Deng}}},\ }\bibfield  {title} {\enquote {\bibinfo {title} {Short-range
  correlations in percolation at criticality},}\ }\href {\doibase
  10.1103/PhysRevE.90.042106} {\bibfield  {journal} {\bibinfo  {journal} {Phys.
  Rev. E}\ }\textbf {\bibinfo {volume} {90}},\ \bibinfo {pages} {042106}
  (\bibinfo {year} {2014})}\BibitemShut {NoStop}%
\bibitem [{\citenamefont {Makse}\ \emph
  {et~al.}(1995{\natexlab{b}})\citenamefont {Makse}, \citenamefont {Havlin},\
  and\ \citenamefont {Stanley}}]{Makse1995a}%
  \BibitemOpen
  \bibfield  {author} {\bibinfo {author} {\bibfnamefont {H.~A.}\ \bibnamefont
  {Makse}}, \bibinfo {author} {\bibfnamefont {S.}~\bibnamefont {Havlin}}, \
  and\ \bibinfo {author} {\bibfnamefont {H.~E.}\ \bibnamefont {Stanley}},\
  }\bibfield  {title} {\enquote {\bibinfo {title} {Modelling urban growth
  patterns},}\ }\href {\doibase 10.1038/377608a0} {\bibfield  {journal}
  {\bibinfo  {journal} {Nature}\ }\textbf {\bibinfo {volume} {377}},\ \bibinfo
  {pages} {608} (\bibinfo {year} {1995}{\natexlab{b}})}\BibitemShut {NoStop}%
\bibitem [{\citenamefont {Stella}\ and\ \citenamefont
  {Vanderzande}(1989)}]{Stella1989}%
  \BibitemOpen
  \bibfield  {author} {\bibinfo {author} {\bibfnamefont {A.~L.}\ \bibnamefont
  {Stella}}\ and\ \bibinfo {author} {\bibfnamefont {C.}~\bibnamefont
  {Vanderzande}},\ }\bibfield  {title} {\enquote {\bibinfo {title} {Scaling and
  fractal dimension of {I}sing clusters at the $d=2$ critical point},}\
  }\href@noop {} {\bibfield  {journal} {\bibinfo  {journal} {Phys. Rev. Lett.}\
  }\textbf {\bibinfo {volume} {62}},\ \bibinfo {pages} {1067} (\bibinfo {year}
  {1989})}\BibitemShut {NoStop}%
\bibitem [{\citenamefont {Duplantier}\ and\ \citenamefont
  {Saleur}(1989)}]{Duplantier1989}%
  \BibitemOpen
  \bibfield  {author} {\bibinfo {author} {\bibfnamefont {B.}~\bibnamefont
  {Duplantier}}\ and\ \bibinfo {author} {\bibfnamefont {H.}~\bibnamefont
  {Saleur}},\ }\bibfield  {title} {\enquote {\bibinfo {title} {Exact fractal
  dimension of 2{D} {I}sing clusters},}\ }\href@noop {} {\bibfield  {journal}
  {\bibinfo  {journal} {Phys. Rev. Lett.}\ }\textbf {\bibinfo {volume} {63}},\
  \bibinfo {pages} {2536} (\bibinfo {year} {1989})}\BibitemShut {NoStop}%
\bibitem [{\citenamefont {Cardy}(1996)}]{Cardy1996}%
  \BibitemOpen
  \bibfield  {author} {\bibinfo {author} {\bibfnamefont {J.}~\bibnamefont
  {Cardy}},\ }\href@noop {} {\emph {\bibinfo {title} {Scaling and
  Renormalization in Statistical Physics}}}\ (\bibinfo  {publisher}
  {{Cambridge} {University} {Press}},\ \bibinfo {address} {Cambridge},\
  \bibinfo {year} {1996})\BibitemShut {NoStop}%
\bibitem [{\citenamefont {Zinn-Justin}(2007)}]{Zinn-Justin2007}%
  \BibitemOpen
  \bibfield  {author} {\bibinfo {author} {\bibfnamefont {J.}~\bibnamefont
  {Zinn-Justin}},\ }\href@noop {} {\emph {\bibinfo {title} {Phase Transitions
  and Renormalization Group}}}\ (\bibinfo  {publisher} {{Oxford} {University}
  {Press}},\ \bibinfo {address} {New York},\ \bibinfo {year}
  {2007})\BibitemShut {NoStop}%
\bibitem [{\citenamefont {Ballesteros}\ \emph {et~al.}(1999)\citenamefont
  {Ballesteros}, \citenamefont {Fern{\`a}ndez}, \citenamefont
  {Mart{\'i}n-Mayor}, \citenamefont {Mu{\~n}oz~Sudupe}, \citenamefont
  {Parisi},\ and\ \citenamefont {Ruiz-Lorenzo}}]{Ballesteros1999a}%
  \BibitemOpen
  \bibfield  {author} {\bibinfo {author} {\bibfnamefont {H.~G.}\ \bibnamefont
  {Ballesteros}}, \bibinfo {author} {\bibfnamefont {L.~A.}\ \bibnamefont
  {Fern{\`a}ndez}}, \bibinfo {author} {\bibfnamefont {V.}~\bibnamefont
  {Mart{\'i}n-Mayor}}, \bibinfo {author} {\bibfnamefont {A.}~\bibnamefont
  {Mu{\~n}oz~Sudupe}}, \bibinfo {author} {\bibfnamefont {G.}~\bibnamefont
  {Parisi}}, \ and\ \bibinfo {author} {\bibfnamefont {J.~J.}\ \bibnamefont
  {Ruiz-Lorenzo}},\ }\bibfield  {title} {\enquote {\bibinfo {title} {Scaling
  corrections: site percolation and {I}sing model in three dimensions},}\
  }\href {http://stacks.iop.org/0305-4470/32/i=1/a=004} {\bibfield  {journal}
  {\bibinfo  {journal} {J. Phys. A: Math. Gen.}\ }\textbf {\bibinfo {volume}
  {32}},\ \bibinfo {pages} {1} (\bibinfo {year} {1999})}\BibitemShut {NoStop}%
\bibitem [{Note2()}]{Note2}%
  \BibitemOpen
  \bibinfo {note} {Note that there is a typo in the argument of the Euler gamma
  function in Ref.~\cite {Makse1996}}\BibitemShut {NoStop}%
\end{thebibliography}%

\end{document}